\documentclass[aps,prd,reprint, superscriptaddress]{revtex4-1}

\usepackage{amsmath,amssymb, amsthm,amstext}
\usepackage{natbib}
\usepackage{graphicx}
\usepackage{color}
\usepackage{array, enumerate}
\usepackage{bm}
\usepackage{multirow}
\usepackage[breaklinks,colorlinks,citecolor=blue]{hyperref}
\usepackage{braket}
\usepackage{txfonts}

\newcommand{\pder}[2][]{\frac{\partial#1}{\partial#2}}
\newcommand{\hbm}[1]{\hat{\bm{#1}}}
\newcommand{\pr}[1]{\left(#1\right)}
\newcommand{\ps}[1]{\left[#1\right]}

\def\be{\begin{equation}}
\def\ee{\end{equation}}

\def\mnras{MNRAS}
\def\aap{A\&A}
\def\apjl{Astrophys.J.Lett.}
\def\aj{Astronomical Journal}

\begin{document}
\title{Mass-gap extreme mass ratio inspirals}
\author{Zhen Pan}
\email{zpan@perimeterinstitute.ca}
\affiliation{Perimeter Institute for Theoretical Physics, Ontario, N2L 2Y5, Canada}
\author{Zhenwei Lyu}
\affiliation{Perimeter Institute for Theoretical Physics, Ontario, N2L 2Y5, Canada}
\affiliation{University of Guelph, Guelph, Ontario N1G 2W1, Canada}
\author{Huan Yang}
\email{hyang@perimeterinstitute.ca}
\affiliation{Perimeter Institute for Theoretical Physics, Ontario, N2L 2Y5, Canada}
\affiliation{University of Guelph, Guelph, Ontario N1G 2W1, Canada}
\begin{abstract}
In  this work, we propose a new subclass of extreme-mass-ratio-inspirals (EMRIs):  mass-gap EMRIs,  consisting of a compact object in the lower mass gap $\sim (2.5-5) M_\odot$
and a massive black hole (MBH). The mass-gap object (MGO) may be  a primordial black hole or produced from a  delayed  supernova
explosion.
We calculate the formation rate of mass-gap EMRIs
in both the (dry) loss-cone channel and the (wet) active galactic nucleus disk channel by solving Fokker-Planck-type equations for the phase-space distribution.
In the dry channel, the mass-gap EMRI rate is strongly suppressed compared to the EMRI rate of stellar-mass black holes (sBHs) as a result of mass segregation effect. In the wet channel, the suppression is roughly equal to the mass ratio of sBHs over MGOs, because the migration speed of a compact object in an active galactic nucleus disk is proportional to its mass. We find that the wet channel is much more promising to produce
mass-gap EMRIs observable by spaceborne gravitation wave detectors.
(Non-)detection of mass-gap EMRIs may  be used to distinguish different supernova explosion mechanisms
and constrain the  abundance of primordial black holes around MBHs.
\end{abstract}
\maketitle

\section{Introduction}

Observations of Galactic X-ray binaries have indicated a dearth of compact objects around $\sim (2.5-5) M_\odot$ in the mass spectrum \cite[e.g.,][]{Bailyn1998,Ozel2010,Ozel2012}.
Whether this mass gap is a result of observational selection effects or underlying
supernova (SN) explosion mechanism has been an open question for more than a decade \cite{Ozel2010,Kreidberg2012,Belczynski2012,Fryer2012}.
More recently new events detected with gravitational waves (GWs) and time-domain astronomy suggest
the existence of mass-gap objects (MGOs) in compact object binaries and in Galactic non-interacting binaries
- the compact objects of $\sim (2.6-2.8) M_\odot$ in
compact binary coalescence events GW190814 and GW200210  \cite{LVC190814,LVC2021},
a compact object of $\sim 3.3 M_\odot$ as a non-interacting companion  of a giant \cite{Thompson2019},
and a compact object of $\sim 3.0 M_\odot$ as a non-interacting companion of a red giant \cite{Jayasinghe2021}
(see \cite{Giesers2018,Giesers2019,Wyrzykowski2020} for candidate MGOs in non-interacting binaries
and as dark lens in the Milky Way).
These identified MGOs provide evidences of a population of compact objects lying in the mass gap, or even a more extreme possibility that the mass gap itself does not exist. A natural question is that,
if indeed a population of MGOs is present, what should be their origin?
One viable option  is MGOs are born in delayed SN explosions \cite{Fryer2012}, which also provides an explanation to the merger rate of GW190814-like events \cite{Zevin2020,Drozda2020}. A more exotic possibility is that  MGOs are primordial black holes (PBHs) \cite{Carr2019,Yang:2017gfb,Clesse2020,Jedamzik2021}, which have been intensively discussed in the context of compact binary mergers detected by LIGO/Virgo. With the upgrading of LIGO/Virgo and the coming era of third-generation detectors, more GW190814-like events are
expected to be detected. However,
as there is already a large number of proposed formation channels \cite[e.g.,][]{Safarzadeh2020,Fragione2020,Yang2020,Rastello2020,
Broadhurst2020,Hamers2021,Lu2021,Liu2021,Tagawa2021b,Arca2021}, many of which are still subject to large theoretical uncertainties (see \cite{Mandel2021} for a recent review),
it is unclear whether we will be able to nail down the the origin of MGOs with only the detection of stellar-mass binaries.

To better answer these questions, we investigate the possiblity that MGOs appear as components of some extreme mass ratio inspirals (EMRIs), to form mass-gap EMRIs. We propose that future detections of EMRIs
and especially mass-gap EMRIs by spaceborne GW detectors, may be a useful probe for the origins of MGOs,  thanks to the limited number of EMRI formation channels and distinct signatures of EMRI sources from different channels.
There are two main channels of EMRI formation: the dry loss-cone channel
\cite{Hopman2005,Preto2010, Bar-Or2016,
Amaro2011,Amaro2018,Amaro2020,Babak2017,Gair2017,Fan2020} and the wet active galactic nucleus (AGN) disk channel \cite{Levin2003,Levin2007, Sigl2007,Tagawa2020,Pan2021a,Pan2021b}
(other processes involving tidal disruption or tidal
capture of binary sBHs, or tidal stripping of giant stars \cite{Miller2005,Chen2018,Wang2019,Raveh2021}
may also contribute to EMRI formation and Naoz et al.\cite{Naoz2022} pointed out  EMRI formation in MBH binaries
is more efficient than around single MBHs).
In the dry channel, a stellar-mass black hole (sBH) is scattered by stars in the nuclear stellar cluster and gravitationally captued by the MBH. There are two relevant timescales in the long-term dynamics:  the GW emission timescale $t_{\rm gw}$ on which the sBH orbit shrink,
and the diffusion timescale $t_J$ on which the orbital angular momentum changes by successive scatterings
\cite{Hopman2005,Amaro2011,Zwick2020,Zwick2021,Vzquez2021}. If GW emission is more efficient with $t_{\rm gw} < t_J$,
the sBH becomes a stable EMRI with continually shrinking orbit until the final coalescence with the MBH.
If scatterings are more efficient, i.e., $t_J < t_{\rm gw}$, the sBH is randomly scattered towards or away from the MBH.
SBHs that are scattered into the MBH without losing much energy via GW emission are called prompt infalls \cite{Hopman2005}. As the GW emission timescale is much shorter for eccentric orbiters ($e\rightarrow 1$), EMRIs in the dry channel are highly eccentric at formation.

A fraction $\mathcal O(10^{-2}-10^{-1})$ of MBHs in the universe (referred as AGNs) are actively accreting gas with an accretion disk \cite{Galametz2009,Macuga2019}. The presence of an accretion disk introduces new interactions affecting the motion
of sBHs in the stellar cluster. For a sBH embedded in the AGN disk, its periodic motion produces density waves
\cite{Goldreich1979,Goldreich1980,Tanaka2002,Tanaka2004} that
in turn drive the sBH to migrate inward, damp its orbital eccentricity and its inclination w.r.t. the disk plane.
For a sBH on a highly inclined orbit, the effects of density waves becomes subdominant to dynamical friction
\cite{Arzamasskiy2018,Zhu2019} arising from the relative motion of the sBH and the surrounding gas as it passes through the disk.
As a result, sBHs are first captured onto the disk driven by dynamical fraction and density waves, and
then migrate inward driven by the density waves upon reaching the vicinity of the MBH, where GW emission become dominant.
We refer this type of EMRIs as wet EMRIs. Because  density waves are very efficient at damping out the eccentricity, wet EMRIs are essentially circular.
Dry and wet EMRIs can be easily distinguished from each other by measuring their eccentricities using  spaceborne gravitational wave detectors \cite{Babak2017,Gair2017,Fan2020}. In addition,  the imprints of an accretion disk on the EMRI waveform may also be detectable \cite{Kocsis2011,Yunes2011}.

In the dry channel, sBHs are generally closer to the central MBH than MGOs due to the mass segregation, so that
their EMRI rate should be larger.
In the wet channel, the EMRI formation rates of both sBHs and MGOs mainly depend on
the capture rate onto the AGN disk and
the rate of migration along the disk.
Both rates are proportional to the object mass,
therefore the EMRI rate of MGOs is lower than that of sBHs by a factor of their mass ratio at most. We find the
wet EMRI rate per AGN  is generally higher than the dry EMRI rate per MBH by $\mathcal O (10^1-10^3)$
for sBHs, and by $\mathcal O (10^3-10^4)$ for MGOs.
 Taking into account the AGN fraction $f_{\rm AGN}=\mathcal O(10^{-2}-10^{-1})$,
the wet channel turns out to be primary way in producing mass-gap EMRIs.
If there is no gap in the mass spectrum of supernova remnants
and roughly equal number of MGOs and sBHs are born in SN explosions,
we expect LISA to detect $\lesssim 1$ dry mass-gap EMRIs per year,
and $\mathcal O(1-10^2)\times (f_{\rm AGN}/1\%)$ wet mass-gap EMRIs per year.
In addition, future detections of dry and wet mass-gap EMRIs have interesting implications of MGO formation.
An excess of mass-gap EMRI detection could be a signature of MGOs of more  exotic origin, e.g., PBHs.
The relative fraction of mass-gap EMRIs to sBH EMRIs is a sensitive probe for the
mass spectrum of SN remnants and therefore the SN explosion mechanism.

The remaining part of this paper is organized as follows.
In Sec.~\ref{sec:lc} and \ref{sec:disk},
we introduce the two EMRI formation channels and the formation rates
of MGOs and sBHs in these two channels.
In Sec.~\ref{sec:detect}, we forecast the detection prospects of these EMRIs by LISA.
We summarize this work in Sec.~\ref{sec:summary}.
Throughout this paper, we use geometrical units $G=c=1$,
and assume a flat $\Lambda$CDM cosmology with $\Omega_{\rm m}=0.307,\Omega_{\rm \Lambda}=1-\Omega_{\rm m}$ and $H_0=67.7$ km/s/Mpc ($h=0.677$).

\section{Dry loss-cone channel}\label{sec:lc}
In this section, we will first briefly introduce the loss-cone mechanism along with the Fokker-Planck equation governing the evolution of stars in a stellar cluster around a MBH. After that we calculate the EMRI formation rates of both
sBHs and MGOs via the loss-cone mechanism.

\subsection{Loss-cone mechanism}
Consider a star orbiting around a MBH, with specific binding energy $E:=\phi-v^2/2$ and
specific angular momentum $J$, where $\phi(r)$ is the (positive) gravitational potential and $v^2/2$
is the kinetic energy. Its orbital motion is affected by two main effects:
GW emission which shrinks the orbit on a timescale $t_{\rm gw}$
and gravitational scatterings by other stars in the stellar cluster
which changes the orbital angular momentum by order of unity on a timescale $t_J$.
For a star on a tight and eccentric orbit where the GW emission is more efficient with $t_{\rm gw} < t_J$, the orbit is stable against random scatterings and the star becomes an stable EMRI \cite{Hopman2005}.
On the other hand, for a star on a wide and/or circular orbit
where the GW emission is less efficient with $t_{\rm gw} > t_J$,
the star is expected to be scattered into a random direction: away from, towards or even directly into the central MBH.

In the phase space, a region of low angular momentum $J < J_{\rm lc}(E)$ is usually referred as
the loss cone, where a star ususally promptly falls into the MBH within one orbital period
$P(E)$ if its angular momentum is not altered much by gravitational scatterings.
As a result, the loss cone region satisfying $P(E) < t_J$ is unpopulated (empty regime) and
the loss cone region satisfying $P(E) > t_J$ is populated (full regime).
For the problem we are investigating, relevant orbits are of low energy
($E\approx 0$)  with semi-major axis length $a \gg M_\bullet$ and the boundary of the loss cone is
\cite{Cutler1994}
\be
J_{\rm lc} (E\approx 0) = 4 M_\bullet \ .
\ee

\subsection{Fokker-Planck equation}
Statistical properties of stars can be described by their distribution
functions $f_i(t,\vec r, \vec v)$ in the $(\vec r, \vec v)$ phase space, where $i$ labels different star species.
Following Refs.~\cite{Cohn1978,Cohn1979}, we approximate the distribution functions as $f_i\approx f_i(t, E, R)$,
where  $R:=J^2/J_c^2(E)$ is the normalized orbital angular momentum with
$J_c(E)$ being the maximum orbital angular momentum of a star with energy $E$.
In order to relate the distribution function $f(E,R)$ to the number density $n(r)$, and derive the Fokker-Planck equation,
it is necessary to understand the  properties of star orbits in given potential field $\phi(r)$,
for which we summarize  as follows \cite{Cohn1979}. The  definition of energy suggests that
\be
2(\phi-E) = v^2 = v_t^2 + v_r^2 = \frac{J^2}{r^2} + v_r^2\ ,
\ee
where $v_t$ and $v_r$ are the tangential velocity and the radial velocity respectively.
For a circular orbit of energy $E$, its orbit radius $r_c(E)$ and angular momentum $J_c(E)$
are determined by
\be
\begin{aligned}
  J_c^2(E) = -r_c^3\phi'(r_c)\ , \\
  2(\phi(r_c)-E) = \frac{J_c^2}{r_c^2}\ .
\end{aligned}
\ee
For a general non-circular orbit with parameters $(E,R)$,
its turning points (apsis/periapsis) $r_\pm$ are determined by
\be
2(\phi(r_\pm) -E) = \frac{J^2}{r_\pm^2}\ ,
\ee
and its orbit period $P(E,R)$ is
\be
P(E,R) = 2\int_{r_-}^{r_+} \frac{dr}{v_r}\ .
\ee
Defining the particle number density in the $(E,R)$ phase space as $N(E,R)$,
with $N(E,R)dEdR:=\int_{r_-}^{r_+} d^3rd^3v f(E,R)$,
we have \cite{Cohn1978,Cohn1979}
\be\label{eq:C}
\begin{aligned}
  N(E,R)
  &= 4\pi^2 P(E,R) J_c^2(E)f(E,R) \\
  :&=\mathcal C(E,R) f(E,R)\ .
\end{aligned}
\ee
The position-space particle number density $n(r)$ can be expressed by the distribution function $f(E,R)$ by \cite{Cohn1979}
\be
n(r) = \frac{2\pi}{r^2}\int_0^{\phi(r)} dE J_c^2(E) \int_0^{R_{\rm max}} \frac{dR}{v_r} f(E,R)\ ,
\ee
where $R_{\rm max}(r, E)= 2r^2(\phi(r)-E)/J_c^2(E)$, and $v_r(r, E,R) = 2(\phi-E)-J^2/r^2 = (R_{\rm max}-R)J_c^2(E)/r^2$.
In the case of thermal distribution $f=f(E)$, the above equation simplifies as \cite{Chernoff1990}
\be
n(r) = 4\pi\int_0^{\phi(r)} dE\sqrt{2(\phi(r)-E)} f(E)\ .
\ee

With all these orbital properties, the Fokker-Planck equation governing the phase space evolution  is written in the form of \cite{Cohn1978,Cohn1979, Shapiro1978,Binney1987, Bar-Or2016}
\be\label{eq:FP}
  \mathcal C\frac{\partial f}{\partial t}
  = - \frac{\partial}{\partial E} F_E
  - \frac{\partial}{\partial R}F_R \ ,
\ee
with $\mathcal C=4\pi^2 J_c^2 P$ being the weight function defined in Eq.~(\ref{eq:C}) and $F_{E,R}$ being the flux in the $E/R$ direction:
\be\label{eq:flux}
\begin{aligned}
  -F_E &= \mathcal C \pr{D_{EE}\frac{\partial f}{\partial E} + D_{ER}\frac{\partial f}{\partial R} + D_E f}\ ,\\
  -F_R &= \mathcal C \pr{D_{RR}\frac{\partial f}{\partial R} + D_{ER}\frac{\partial f}{\partial E} + D_R f}\ .
\end{aligned}
\ee
The diffusion and advection coefficients are functions of the distribution functions and are derived in Appendix~\ref{apa}.

\begin{figure*}
\includegraphics[scale=0.5]{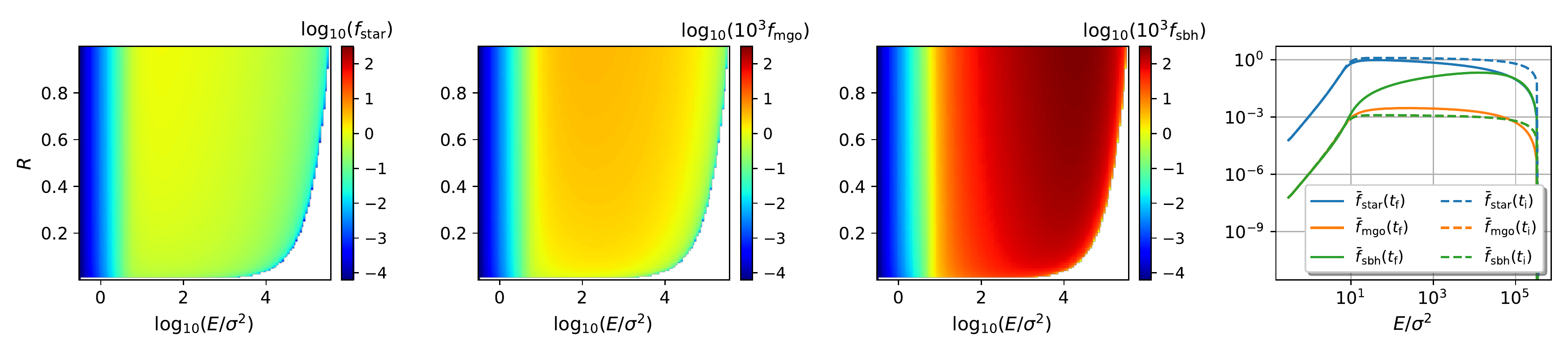}
\caption{\label{fig:lc}  In the fiducial stellar cluster around a MBH with $M_\bullet=4\times10^6 M_\odot$, distribution functions $f_i(E,R)$ ($i={\rm star, mgo, sbh}$)
at $t_{\rm f}= 2$ Gyr are shown in the first 3 panels. The 4th panel shows $R$-integrated distribution functions $\bar f_i$ at $t_{\rm i}=0$ (dashed lines) and at $t_{\rm f}=2$ Gyr (solid lines). All the distribution functions are shown in units of $10^5{\rm pc}^{-3}/(2\pi\sigma^2)^{3/2}$.}
\end{figure*}

Given initial condition $f(t=0, E, R)$, we evolve the cluster according to the Fokker-Planck equation (\ref{eq:FP})
subject to following boundary conditions.
On the $E\rightarrow 0$ boundary, we fix the distributions to their initial values, i.e.,
\be
f(t,E,R)|_{E\rightarrow 0} = f(t=0,E,R)|_{E\rightarrow 0}\ ,
\ee
considering the long relaxation timescale there.
On the $R=1$ boundary, the flux in the $R$ direction should vanish,
\be
F_{R}|_{R \rightarrow 1} = 0 \ .
\ee
On the loss cone boundary $R=R_{\rm lc}(E):=J_{\rm lc}^2/J_c^2(E)$,
there are two different regimes: full loss cone regime where
\be
y_{\rm lc}:= \frac{R_{\rm lc}}{(D_{RR}/R)_{R\rightarrow0} P} < 1\ ,
\ee
and empty regime where $y_{\rm lc} > 1$. In the empty regime, stars are  expected to fall into the MBH within one
orbital period $P$ (that's why the phase space is empty). In the full regime,
stars are in general scattered into/out of the loss cone mulitiple times within one orbital period,
therefore the phase space is full of stars and the rate of stars falling into the MBH is low.
Quantitatively, the flux in the $R$ direction was obtained in Ref.~\cite{Cohn1978} as
\be
-\frac{F_R}{\mathcal C} = \left( \frac{D_{RR}}{R}\right)\Bigg|_{R\rightarrow 0}  \frac{f(R_0)}{\ln(R_0/R_{\rm lc}) + \mathcal F(y_{\rm lc})}\ ,
\ee
where $R_0$ is any small $R$ in the range of $R_{\rm lc}\le R \ll 1$, $\mathcal F(y_{\rm lc})\sim 1/y_{\rm lc}$ for $y_{\rm lc}\lesssim 1$ and $\mathcal F(y_{\rm lc})\simeq 0.824y_{\rm lc}^{-1/2}$ for $y_{\rm lc}\gtrsim 1$.
At $R_0=R_{\rm lc}$, the above equation simplifies as  $F_R(R_{\rm lc}) = 0$ in the full regime,
and $f(R_{\rm lc})= 0$ in the empty regime.

As a result, the EMRI rate and the promp infall rate per MBH via loss cone is given by
\be\label{eq:emri}
\begin{aligned}
  \Gamma_{\rm emri, lc} &= \int_{E_{\rm gw}}^{+\infty}  \vec F \cdot d\vec l\ , \\
  \Gamma_{\rm infl, lc} &= \int_{\sigma^2}^{E_{\rm gw}}  \vec F \cdot d\vec l\ ,
\end{aligned}
\ee
where  $\vec F = (F_E, F_R)$, $d\vec l = (dE, dR)$ is the line element along the boundary of the loss cone,
and $E_{\rm gw}$ is the critical energy where $t_{\rm gw} = t_J$.
To calculate the GW emission timescale $t_{\rm gw}$, we use a recently corrected version of Peters' time-scale that accounts for eccentricity evolution and post-Newtonian corrections \cite{Peters1964,Zwick2020,Zwick2021,Vzquez2021}, with
\be\label{eq:tgw}
t_{\rm gw} = \frac{5 a^4}{256 M_\bullet^2 m}\frac{(1-e^2)^{7/2}}{1+\frac{73}{24}e^2+\frac{37}{96}e^4}
8^{1-\sqrt{1-e}}e^{\frac{5M_\bullet}{a(1-e)}}\ ,
\ee
where $m$ is the mass of the star orbiting around the MBH,
$a$ and $e$ are the orbital semi-major axis and the eccentricity, respectively.
For calculating the diffusion timescale $t_J$ in the $J$-direction, we use the approximation \cite{Hopman2005}
\be\label{eq:tJ}
t_J \approx \frac{J^2}{J_c^2(E)} t_E(E, R) = \frac{J^2}{J_c^2(E)}\frac{E^2}{2D_{EE}(E,R=0)}\ .
\ee

\subsection{EMRI rate and prompt infall rate}

\begin{figure}
\includegraphics[scale=0.7]{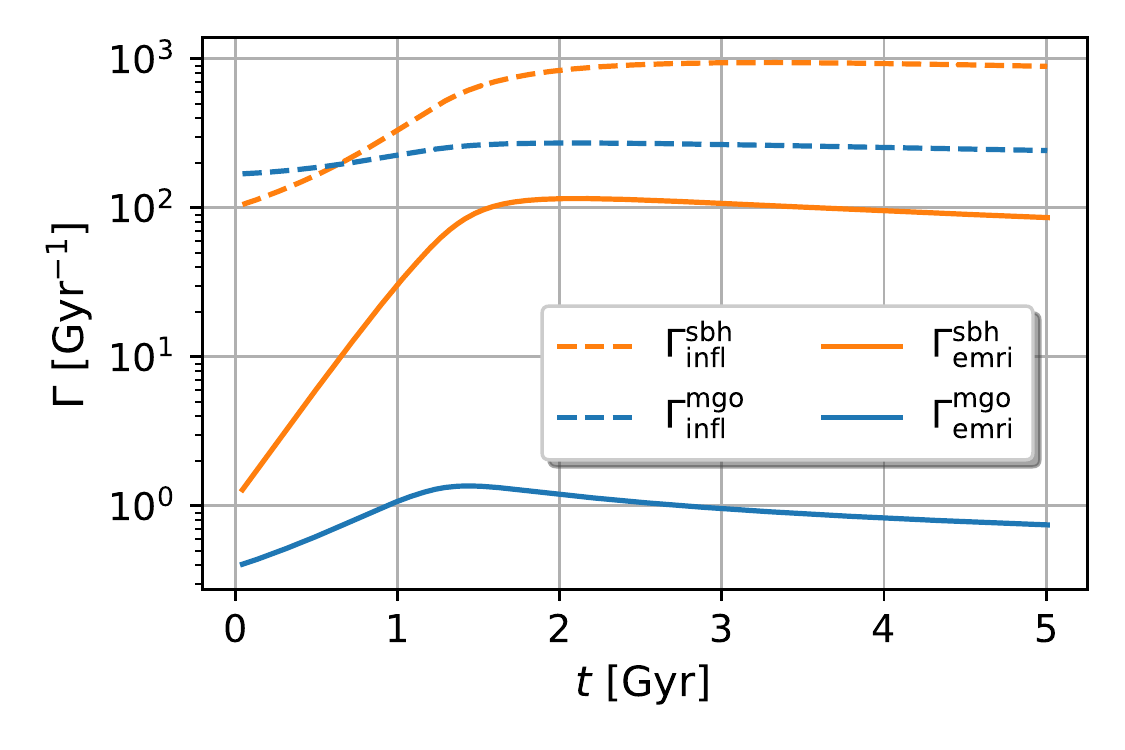}
\includegraphics[scale=0.7]{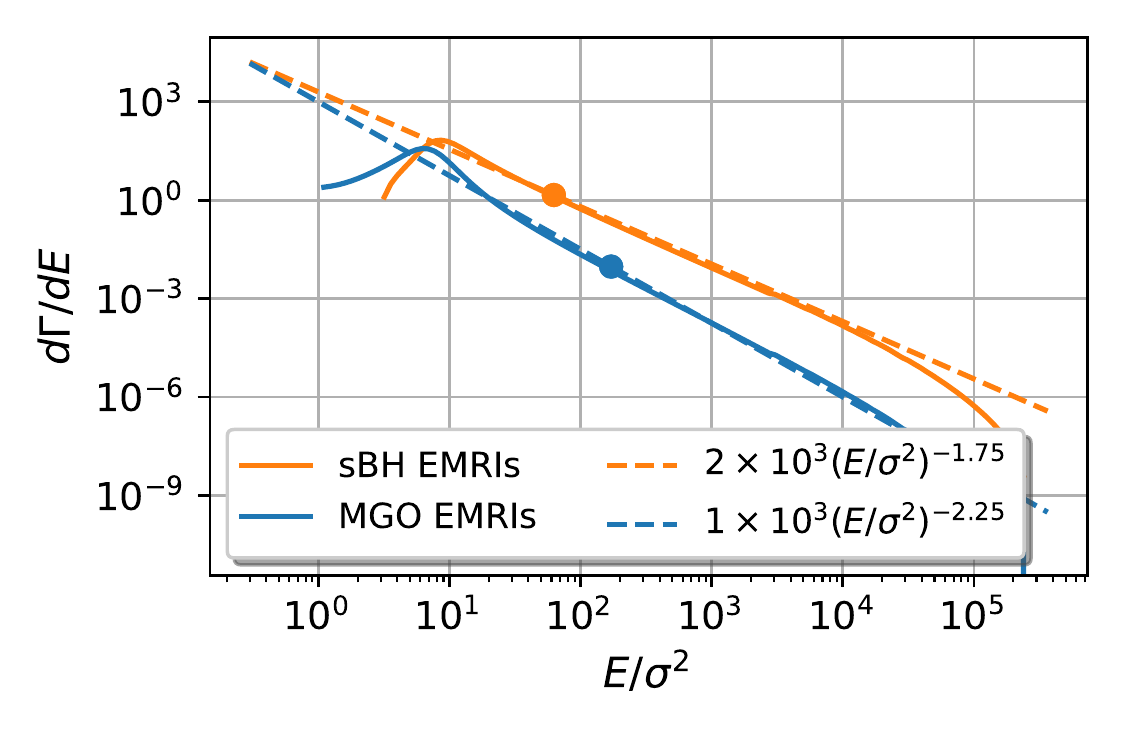}
\caption{\label{fig:Gamma_lc} Upper panel: EMRI rates $\Gamma_{\rm emri, lc}$ and prompt infall rates  $\Gamma_{\rm infl,lc}$ of both sBHs and mass-gap objects in the loss-cone channel.
Lower panel: differential rates $d\Gamma/dE$ at $t=$ 2 Gyr, and the two dots are the critical energy
 $E_{\rm gw}^{\rm mgo, sbh}$.}
\end{figure}

We initialize the system with Tremaine's MBH+stellar cluster model \cite{Tremaine1994,Dehnen1993}, assuming 3 star species in the stellar cluster: stars with mass $m_{\rm star}$, mass-gap objects with mass  $m_{\rm mgo}$ and heavy sBHs with mass $m_{\rm sbh}$.
The total star/mgo/sBH mass in the cluster
are $M_{\rm star}$, $M_{\rm mgo}$ and $M_{\rm sbh}$, respectively.
Their number densities  in the Tremaine's cluster model are specified by
\be\label{eq:numden}
\begin{aligned}
  n_{\rm star}(r)& = \frac{M_{\rm star}}{m_{\rm star}}\frac{3-\gamma}{4\pi}\frac{r_a}{r^\gamma (r+r_a)^{4-\gamma}}, \\
 n_i(r) &= \delta_i \times n_{\rm star}(r)\ ,
\end{aligned}
\ee
with $i$ the index labelling different star species, $r_a$ the density transition radius, $\gamma$  the density scaling power index, and $\delta_i$  the abundance of species $i$ relative to stars.

As an example, we initialize a stellar cluster with three different star species with $m_i = (1, 3, 10) M_\odot$, $\delta_i=(1, 10^{-3}, 10^{-3})$ around a MBH with $M_\bullet=4\times 10^6 M_\odot$. Note if there is no gap
in the  mass spectrum of SN remnants with the power-law mass spectrum
$dN/dm \propto m^{-2.35}$ \cite{Perna2019}
holding in the whole mass range $3 M_\odot \leq m \leq 50 M_\odot$,
we expect nearly equal number of MGOs and sBHs produced in SN explosions.
The total star mass $M_{\rm star}=20 M_\bullet$, the density transition radius $r_a=4r_{\rm h}=4M_\bullet/\sigma^2$ and
the density power index $\gamma=1.5$, where the star velocity dispersion $\sigma$ satisfies the $M_\bullet-\sigma$ relation \cite{Tremaine2002,Gultekin2009}
\be
M_\bullet = 1.53\times 10^6 \pr{\frac{\sigma}{70\ {\rm km/s}}}^{4.24}\ ,
\ee

We evolve the cluster according to the Fokker-Planck equation (\ref{eq:FP}) (see \cite{Pan2021a} for detailed numerical algorithm). In first 3 panels of Fig.~\ref{fig:lc}, we show the distribution functions $f_i(t,E,R)$ at $t=2$ Gyr,
and in the 4th panel, we show the $R$-integrated functions $\bar f_i(E)=\int_0^1 dR f_i(t,E,R)$ at $t=0$ and $t=2$ Gyr, respectively. From the  4th panel, we see sBHs (which are the most massive star component) concentrate around the MBH
as a result of mass segregation,  yielding a large increase in the distribution function for sBHs
at small radii/large binding energy $E$ with time
\cite{Alexander2009,Preto2010,Amaro2011}, while little concentration is found for less massive MGOs or stars.

In the upper panel of Fig.~\ref{fig:Gamma_lc}, we show the EMRI rates and the prompt infall rates of both sBHs and MGOs,
where $\Gamma_{\rm emri}^{\rm mgo}$ is lower than $\Gamma_{\rm emri}^{\rm sbh}$ by a factor of $\mathcal O(10^2)$,
as a result of the stronger mass segregation and shorter GW emission timescale $t_{\rm gw}$  for sBHs.
In contrast, the prompt infall rates are less affected by the mass segregation,
because the prompt infall rate depends on the star density at lower energy ($<E_{\rm gw})$,
while the EMRI rate depends on the star density at higher energy ($>E_{\rm gw}$) (Eq.~\ref{eq:emri}),
and the latter is more sensitive to the mass segregation  (Fig.~\ref{fig:lc}).
As a result, we find the number of prompt infalls per EMRI
$N_{\rm p}:=\Gamma_{\rm infl}/\Gamma_{\rm emri}$ are $N_{\rm p}^{\rm sbh} \approx 10, N_{\rm p}^{\rm mgo} \approx 250$ for the fiducial model.

For an analytic understanding of these results, we re-estimate the number of prompt infalls per EMRI $N_{\rm p}$ using previous analytic formula [Eq.~(17) and (26) in \cite{Hopman2005}],
\be
\begin{aligned}
\Gamma_{\rm emri,lc} = \int_0^{a_{\rm gw}} \frac{4\pi a^2 n(a)}{\ln(R_{\rm lc})t_{\rm rlx}(a)} \ da \sim a_{\rm gw} ^{1.5-2p}\ , \\
\Gamma_{\rm infl,lc} = \int_{a_{\rm gw}}^{a_{\rm max}} \frac{4\pi a^2 n(a)}{\ln(R_{\rm lc})t_{\rm rlx}(a)} \ da \sim a_{\rm max} ^{1.5-2p}\ ,
\end{aligned}
\ee
where $n(a)\sim a^{-1.5-p}$ is the number density, $t_{\rm rlx}(a)\sim a^{p}$ is the local relaxation timescale,
$a_{\rm max}$ is a characteristic radius of the stellar cluster,
and $a_{\rm gw}$ is the critical radius where $t_{\rm gw}=t_J$ [Eqs.(\ref{eq:tgw},\ref{eq:tJ})].
For a single-species cluster filled with stars of mass $m$, the dependence of $a_{\rm gw}$ on mass $m$
and on the MBH mass $M_\bullet$ is \cite{Hopman2005,Naoz2022},
\be\label{eq:agw}
\frac{a_{\rm gw}}{M_\bullet/\sigma^2}\propto m^{\frac{2}{3-2p}} M_\bullet^0 \ .
\ee
For comparison with numerical results, we formulate the above analytic results in the phase space,
with the analytic differential rates
\be
d\Gamma/dE = d\Gamma/da\times da/dE \sim E^{2p-2.5}\ ,
\ee
which can be directly compared wtih the numerical results of the  fiducial model at $t=2$ Gyr (lower panel of Fig.~\ref{fig:Gamma_lc}). From the comparison, we see  the power laws $d\Gamma_{\rm mgo}/dE\sim E^{-2.25}$ and $d\Gamma_{\rm sbh}/dE\sim E^{-1.75}$ are a good approximation at $E\gtrsim 10\sigma^2$, i.e., $2p_{\rm mgo}\approx0.25, 2p_{\rm sbh}\approx0.75$.
In terms of the differential rates, the EMRI rate and the prompt infall rates are written as
\begin{equation}
  \begin{aligned}
    \Gamma_{\rm emri,lc}&= \int_{E_{\rm gw}}^\infty d\Gamma/dE \ dE\ ,\\
      \Gamma_{\rm infl,lc}&= \int_{\sigma^2}^{E_{\rm gw}} d\Gamma/dE \ dE\ ,
  \end{aligned}
\end{equation}
where the critical energy are numerically found as $E_{\rm gw}^{\rm sbh} = 63\sigma^2,
E_{\rm gw}^{\rm mgo}= 171\sigma^2$ and they are consistent with the analytic expectation
$E_{\rm gw}^{\rm sbh}/E_{\rm gw}^{\rm mgo}\approx(m_{\rm sbh}/m_{\rm mgo})^{\frac{2}{3-2p}}$ [Eq.~(\ref{eq:agw})].
 Using the power-law approximations to the differential rates, we have
 \be\label{eq:Np}
 N_{\rm p}\approx (E_{\rm gw}/E_{\rm min})^{1.5-2p},
 \ee
 where $E_{\rm min} \in(1,10)\sigma^2$ is an effective minimum energy (lower panel of Fig.~\ref{fig:Gamma_lc}).
 As a result, we obtain an analytic estimate $N_{\rm p}^{\rm sbh}\in(4,22)$ and
$N_{\rm p}^{\rm mgo}\in (35,600)$.

\begin{figure}
\includegraphics[scale=0.7]{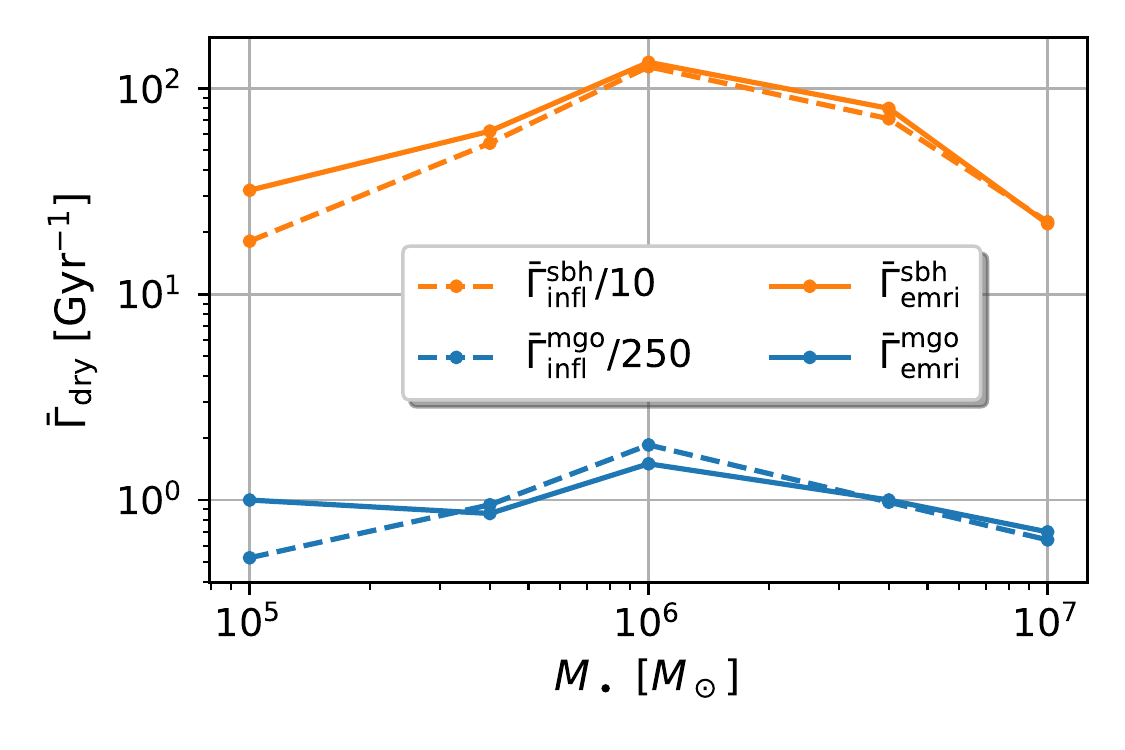}
\caption{\label{fig:Gamma_lc_all} Average EMRI rates $\bar\Gamma_{\rm emri, dry}$ and average prompt infall rates  $\bar\Gamma_{\rm infl,dry}$ of both sBHs and mass-gap objects in the loss-cone channel.}
\end{figure}

Similar to the fiducial model, we initialize the stellar cluster around a MBH with mass in the range of $(10^5, 10^7) M_\odot$, then evolve the system for $T_0=5$ Gyr, and summarize
the time-averaged EMRI rates $\bar\Gamma_{\rm emri}^{\rm sbh, mgo}$ and prompt infall rates $\bar\Gamma_{\rm infl}^{\rm sbh, mgo}$ in Fig.~\ref{fig:Gamma_lc_all}. We find the EMRI rates peak around $M_\bullet=10^6 M_\odot$, because the rates are limited by the longer relaxation timescale of the stellar cluster around a heavier MBH, while the rates are limited by the lower number of sBHs and MGOs in the stellar cluster around a lighter MBH \cite{Pan2021a}. We find the time-averaged EMRI rates are  $\bar\Gamma_{\rm emri}^{\rm sbh}=\mathcal O(10-10^2)\ {\rm Gyr}^{-1}$ and $\bar\Gamma_{\rm emri}^{\rm mgo}=\mathcal O(1)\ {\rm Gyr}^{-1}$. For longer evolution time $T_0$, the rates decrease further because of the depletion of sBHs and MGOs in the stellar cluster.
On average, the number of prompt infalls per EMRI $N_{\rm p}$ are
similar to in the fiducial model with $N_{\rm p}^{\rm sbh}\approx 10, N_{\rm p}^{\rm mgo} \approx 250$,
except $N_{\rm p}(M_\bullet=10^5M_\odot)$ is lower by a factor $\sim 2$.
The nearly independence of $N_{\rm p}$ on the MBH mass $M_\bullet$ comes
from the independence of $a_{\rm gw}$ (or equavilently $E_{\rm gw}$) on $M_\bullet$ [Eq.~(\ref{eq:agw},\ref{eq:Np})], while
$N_{\rm p}^{\rm mgo, sbh}(M_\bullet=10^5M_\odot)$ are lower simply because
MGOs/sBHs around the lighter MBH are of lower number and are quickly depleted via the loss cone,
consequently the critical energy $E_{\rm gw}$
decreases and $N_{\rm p}$ is reduced.

\section{Wet AGN disk channel}\label{sec:disk}
In the presence of an accretion disk around a MBH, the distributions of all different orbiting object species are
affected by the disk. As a result,
the spherical symmetry is broken and the distribution function $f(E,R,\mu)$ generally acquires
dependence on the orbital inclination $\iota$ w.r.t. the disk plane,
where we have defined $\mu:=\cos\iota = \hbm{J}\cdot\hbm{J}_{\rm disk}$,
with $\hbm{J}$ and $\hbm{J}_{\rm disk}$ being the unit direction vectors of
the star orbital angular momentum and the disk angular
momentum, respectively.

For the problem we are considering,
all stars can be conveniently decomposed as a cluster component and a disk component, i.e.,
\be
f(E, R, \mu) \rightarrow f(E, \mu) + g(E)\delta(\mu-1)\delta(R-1)\ ,
\ee
where we have approximate the cluster-component distribution as $R$-independent and
approximate the disk component as circular orbiters lying on the equator with $\iota=0$ ($\mu=1$),
because the orbital eccentricity damping timescale is in general much shorter than the migration timescale
(see subsection~\ref{subsec:star-disk} for details).
With this decomposition, we have number density $n(r,\theta)$
of the cluster-component stars and  surface number density $\Sigma(r)$ of the disk-component stars as
\be
\begin{aligned}
  n(r,\theta) &= 4\pi\int_0^{\phi(r)} dE \sqrt{2(\phi(r)-E)} \bar f(E,\theta)\ ,\\
  \Sigma(r) &=  2\pi^2 r E\sqrt{2(\phi(r)-E)} g(E)\Big|_{r=r_c(E)}\ ,
\end{aligned}
\ee
with
\be
\bar f(E,\theta) = \frac{1}{2\pi}\int_0^{2\pi} d\eta f(E, \mu=\sin\theta\cos\eta)\ ,
\ee
where $\theta$ is the polar angle w.r.t. to the $\hbm{J}_{\rm disk}$ direction.

In the remaining part of this section, we will first summarize the important interactions
between stars and the accretion disk, then incorporate these interactions into the Fokker-Planck equation,
and finally evolve the stellar cluster to calculate the wet EMRI rates.

\subsection{Star-disk interactions}\label{subsec:star-disk}

Interactions of an accretion disk with stars, MGOs and sBHs are similar in aspects of density waves and dynamical friction.
In term of gas accretion onto stars, MGOs and sBHs within the AGN disk, the star size makes a difference.
Compact objects of relatively small sizes only grow mildly within the AGN disk, while
the stellar evolution is expected to be impacted by gas accretion onto stars of much larger sizes \cite{Pan2021c,Jermyn2021}.  In this work, we are not intended to model the star evolution in detail,
and we simply assume no mass change of all the star species during the evolution period.

For illustration purpose, we take sBHs as an example.
As a sBH orbits around the central MBH surrounded by an accretion disk,
its periodic motion generates density waves, which in turn drive the star to migrate inward,
damp its orbital eccentricity $e$ and its orbital inclination $\iota$.
For a highly inclined orbiter, the density wave effects become subdominant with respect to dynamical friction
as it goes through the gas disk.
The two effects (density waves and dynamical friction) together contribute to the advection in $(E, \mu)$ space as
\cite{Arzamasskiy2018,Zhu2019}
\be\label{eq:wave_friction}
\begin{aligned}
  \braket{\Delta \mu}_t^{\rm dsk}
  &= (1-\mu^2)\frac{\iota}{\sin\iota}
  \times {\rm min.}\left\{\frac{0.544}{t_{\rm wav}}, \frac{1.46}{t_{\rm wav}}\frac{h^4}{\iota\sin^3(\iota/2)}\right\}\ ,\\
  \braket{\Delta E}_t^{\rm dsk}
  &= E\times {\rm min.}\left\{\frac{2.7+1.1\alpha_s}{t_{\rm mig,I}}, \frac{8.8}{t_{\rm mig,I}}\frac{h^2}{\sin(\iota)\sin(\iota/2)} \right\}\ ,
\end{aligned}
\ee
with $\braket{\Delta X}_t := \frac{\Delta X}{\Delta t}|_{\Delta t\rightarrow 0}$, $\alpha_s := d\ln \Sigma/d\ln r$ and
\be
t_{\rm mig,I} = \frac{M}{m}\frac{M}{\Sigma r^2} h^2\ , t_{\rm wav} = t_{\rm mig,I} h^2\ ,
\ee
where $m$ is the mass of the orbiter, $M$ is the total mass within radius $r$, $h$ and $\Sigma$ are
disk scale height and the disk surface density \footnote{In the previous work \cite{Pan2021a}, we have approximated the effect of density waves as inclination-independent and have neglected the contribution from dynamical friction. As a result, the previous approximation is an over estimate for highly inclined orbiters.}.

For a sBH captured into the disk, its orbital eccentricity will
be damped by the eccentricity density waves on timescale $t_{\rm wav}$,
which is generally much shorter than all other relevant timescales, including the migration timescale $t_{\rm mig,I}$
and two diffusion timescales, $E^2/D_{EE}$ and $(1-\mu^2)/D_{\mu\mu}$.
As a result, sBHs in the AGN disk are generally moving in circular orbits.
For a sBH embedded in the gas disk, surrounding gas tends to flow towards it
nearly in the radial direction at large distances, in the rest frame of the sBH.
Due to the differential rotation of the gas disk, the inflowing gas generally carries non-zero
angular momentum relative to the sBH, consequently circularizes and forms a disk around the sBH.
The gas inflow rate at the outer boundary is usually super-Eddington and a strong outflow naturally emerges.
As a result, a major part of the inflowing gas may escape as outflow and only the remaining
part is accreted by the sBH (see Ref.~\cite{Pan2021c} and references therein for detailed modeling).
Because of the circularization process, it is reasonable to expect that the outflow
carries minimal net momentum with respect to the sBH, and the momentum carried by the inflow eventually
transfers to the sBH. The head wind contributes to the advection in the $E$-direction as
\be\label{eq:wind}
\braket{\Delta E}_t^{\rm wnd} = \frac{2J}{\dot J_{\rm wnd}}\quad (\textrm{for\ in-disk\ orbiters}) ,
\ee
where $\dot J_{\rm wnd}$ is sBH angular momentum loss rate
due to the head wind (see Ref.~\cite{Pan2021a} for calculation details).

GW emission only becomes important when the orbiter is very close to the MBH
and it drives an advection in the $E$ direction for a circular orbiter as \cite{Peters1964}
\be\label{eq:gw}
\braket{\Delta E}_t^{\rm gw} = \frac{64}{5}\frac{M^2 m}{a^4} E\ .
\ee

\begin{figure}
\includegraphics[scale=0.7]{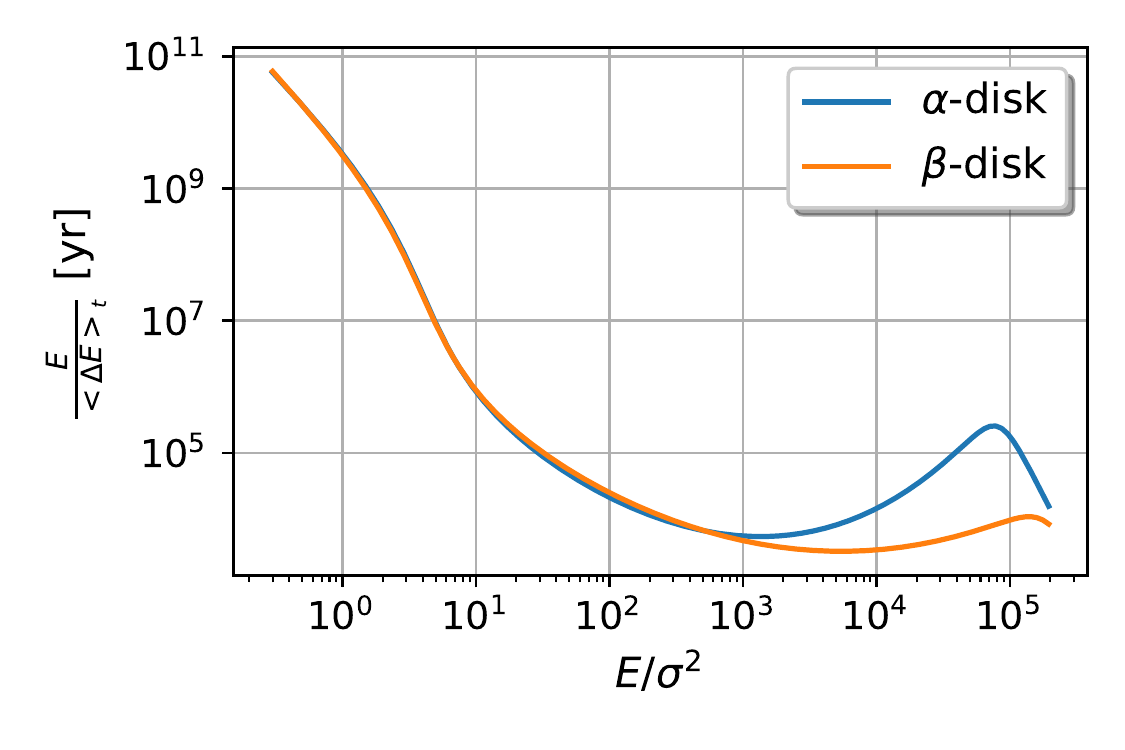}
\caption{\label{fig:tmig} The migration timescales of a $10 M_\odot$
BH embedded in the fiducial $\alpha$ and $\beta$ disks,
respectively, where the GW emission becomes dominant at $E/\sigma^2\sim 10^5$ or equivalently $a\sim 10^2 M_\bullet$.}
\end{figure}

Accretion disk structure of AGNs has not been well understood especially in the outer parts, where both disk heating
mechanism and the angular momentum transport mechanism are not clear. Three commonly used AGN disk models
($\alpha/\beta$ disks \cite{Sirko2003}, and TQM disk proposed by Thompson, Quataert, and Murray \cite{Thompson2005}) have
been numerically solved
and compared in our previous work \cite{Pan2021a}. In this work, we will use $\alpha$ and $\beta$ disks
with accretion rate $\dot M_\bullet = 0.1 \dot M_{\bullet,\rm Edd}$ as fiducial disk models (Fig.~5 in \cite{Pan2021a}).
In Fig.~\ref{fig:tmig}, we show the migration timescales $E/\braket{\Delta E}_t:=E/(\braket{\Delta E}_t^{\rm dsk}+\braket{\Delta E}_t^{\rm wnd}+\braket{\Delta E}_t^{\rm gw})$ of a sBH with $m_{\rm sbh} =10 M_\odot$ embedded in the two fiducial accretion disks.
The two disks only differs in inner parts where the radiation pressure dominates over the gas pressure
and the two migration timescales only differs where $E/\sigma^2\gtrsim 10^3$.
We do not include the TQM disk model in this work because a much more efficient angular momentum transport mechanism is assumed in the TQM disk model, which is inconsistent with the turbulence viscosity driven by magnetorotational instability in inner parts of the accretion disk \cite{Balbus1991,Balbus1998,Martin2019}.
Due to the high efficiecy of the angular momentum transport assumed in the TQM disk model,
TQM disks are in general of lower surface density, therefore longer migration timescale $t_{\rm mig,I}$,
which hinders sBHs and MGOs from migrating to the vicinity of the central MBH and forming EMRIs if $T_{\rm disk} < t_{\rm mig,I}$. If $T_{\rm disk} > t_{\rm mig,I}$, the EMRI rates in TQM disks are similar to those in $\alpha/\beta$-disks  (see \cite{Pan2021a} for details).

\subsection{Fokker-Planck equation}
For cluster-component stars, the Fokker-Planck equation takes the form
\be\label{eq:FP_b}
\mathcal C_\mu \pder[f]{t}
= -\pder{E}F_E- \pder{\mu}F_\mu\ ,
\ee
with flux
\be
\begin{aligned}
  -F_E   &= \mathcal C_\mu \pr{D_{EE}\pder[f]{E}+D_{E\mu}\pder[f]{\mu} + D_E f}\ ,\\
  -F_\mu &=\mathcal C_\mu \pr{D_{\mu\mu}\pder[f]{\mu}+D_{E\mu}\pder[f]{E}+D_\mu f}\ ,
\end{aligned}
\ee
and the weight function
\be
\mathcal C_\mu(E) :=\frac{1}{2}\int_0^1 \mathcal C(E,R) dR \ ,
\ee
where the factor $2$ comes from $\int_{-1}^{1} d\mu$.
All the coefficients of the Fokker-Planck equation (\ref{eq:FP_b}) are contributed by star interactions
with the accretion disk and by scatterings with both the cluster-component stars and the disk-component stars,
where the first only contributes to the advection coefficients as
\be
D_E = -\pr{\braket{\Delta E}_t^{\rm dsk}+\braket{\Delta E}_t^{\rm gw}},\ D_\mu = -\braket{\Delta \mu}_t^{\rm dsk}\ .
\ee
and the latter two contributions are given in Eqs.(\ref{eq:D_cls},\ref{eq:D_dsk}), respectively.

With proper initial conditions, we evolve the system according to Eq.~(\ref{eq:FP_b}) with the following boundary
conditions.
On the $E\rightarrow 0$ boundary, we again fix the distribution, i.e.,
\be
f(t,E,R)|_{E\rightarrow 0} = f(t=0,E,R)|_{E\rightarrow 0}\ .
\ee
On the $E\rightarrow E_{\rm max}$ boundary, where the evolution of distribution function
is dominated by GW emission, we set
\be
F_{E}|_{E \rightarrow E_{\rm max}} =-\mathcal C_\mu D_E f\ .
\ee
On the $\mu=-1$ boundary, we use the zero-flux condition
\be
F_{\mu}|_{\mu \rightarrow -1} = 0 \ .
\ee
On the $\mu=1$ boundary, where the evolution of distribution function is dominated by the inclination damping
arising from the normal density waves, we set
\be
F_{\mu}|_{\mu \rightarrow 1} =-\mathcal C_\mu D_\mu f\ .
\ee

\begin{figure*}
\includegraphics[scale=0.5]{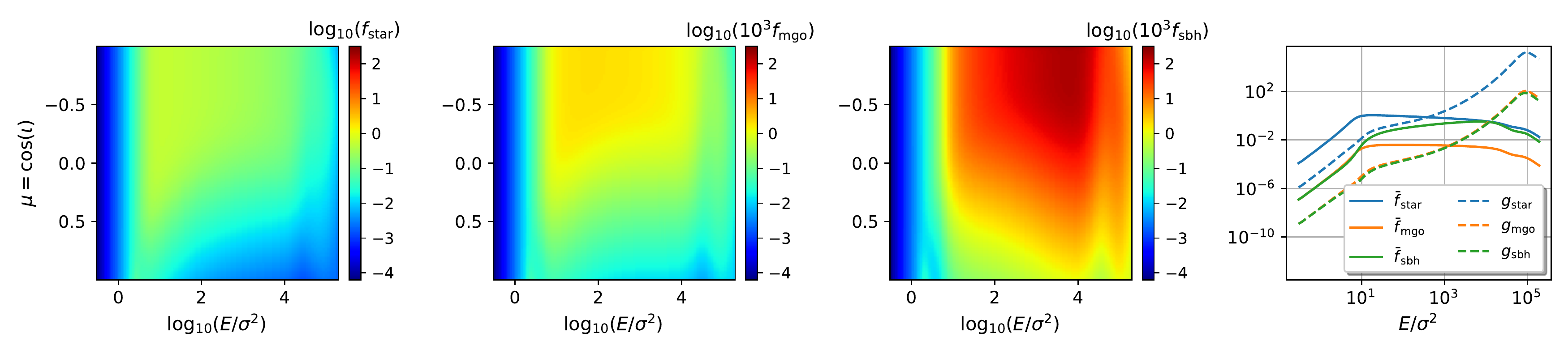}
\caption{\label{fig:disk} In the fiducial stellar cluster around a MBH with $M_\bullet=4\times10^6 M_\odot$, the cluster-component distribution functions $f_i(E,\mu)$ ($i={\rm star, mgo, sbh}$)
at $t= 10^7$ yr are shown in the first 3 panels. The 4th panel shows $\mu$-integrated cluster-component distribution functions $\bar f_i(E)$ (solid lines) and disk-component $g_i(E)$ (dashed lines). All the distribution functions are shown in units of $10^5{\rm pc}^{-3}/(2\pi\sigma^2)^{3/2}$.}
\end{figure*}

For disk-component stars, the Fokker-Planck equation reduces to be 1-dimensional,
\be\label{eq:FP_c}
\mathcal C_\mu \pder[g]{t}
= \pder{E}\ps{ \mathcal C_\mu \pr{D_{EE}\pder[g]{E} + D_E g} }
+ F_\mu(E, \mu=1)\ ,
\ee
with the source term arising form disk-component stars captured by the disk.
We also find the evolution of disk-component stars is dominated by the star-disk interactions and GW emission [Eqs.~(\ref{eq:wave_friction},\ref{eq:wind},\ref{eq:gw})],
so we simply neglect the contributions from scatterings, i.e.,
\be
D_E = -\pr{\braket{\Delta E}_t^{\rm dsk}+\braket{\Delta E}_t^{\rm gw}+\braket{\Delta E}_t^{\rm wnd}}\ , \quad
D_{EE} = 0\ .
\ee
Now Eq.~(\ref{eq:FP_c}) is an first-order differential equation, which requires only one boundary condition,
and we choose it as
\be
g(t,E,R)|_{E\rightarrow 0} = g(t=0,E,R)|_{E\rightarrow 0}\ .
\ee

The wet EMRI rate is determined by the flux in the $E$ direction at the $E_{\rm max}$ boundary, i.e.,
\be
\Gamma_{\rm emri, disk} = -\mathcal C_\mu D_E g |_{E=E_{\rm max}}\ .
\ee
Strictly speaking, the cluster-component contribution  should also be included in addition to the disk-component
contribution. As we will see later, the disk-component constribution at the $E_{\rm max}$ boundary turns out to be dominant.

\begin{figure}
\includegraphics[scale=0.7]{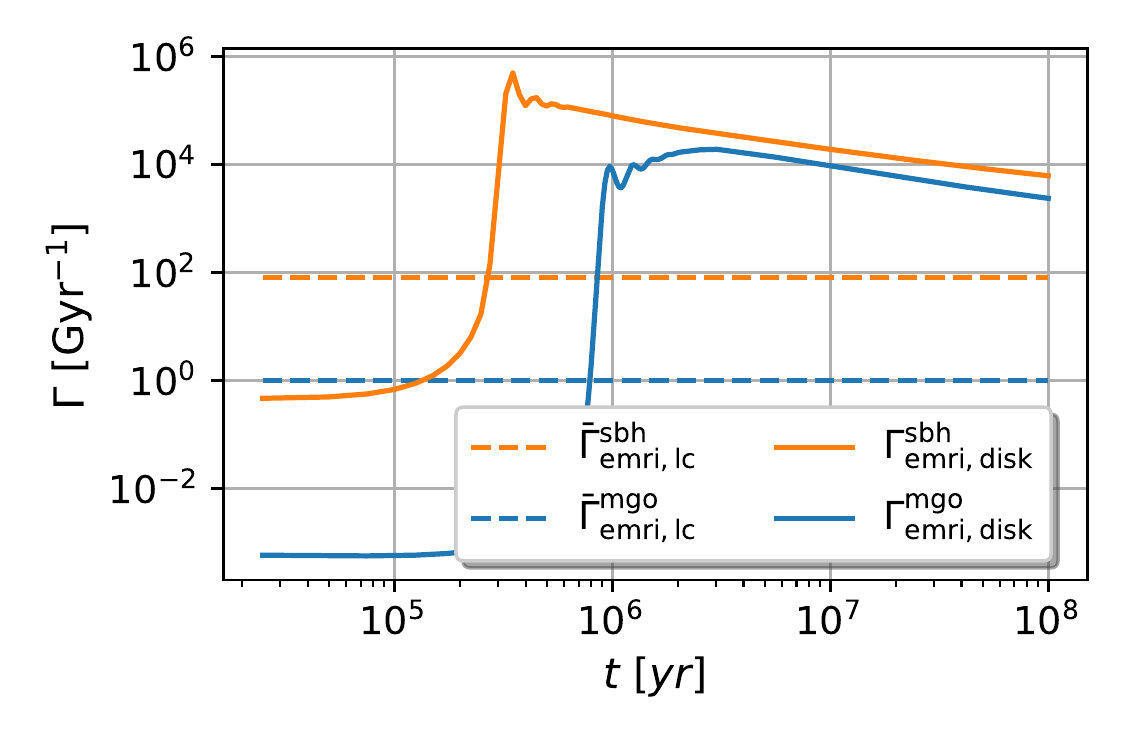}
\caption{\label{fig:Gamma_disk} Wet EMRI rates $\Gamma_{\rm emri, disk}(t)$ of sBHs and MGOs.
For comparision, we also plot the average dry EMRI rates $\bar\Gamma_{\rm emri, lc}$ during the quiet phase  as
 horizontal lines (Fig.~\ref{fig:Gamma_lc_all}).}
\end{figure}

\begin{figure}
\includegraphics[scale=0.7]{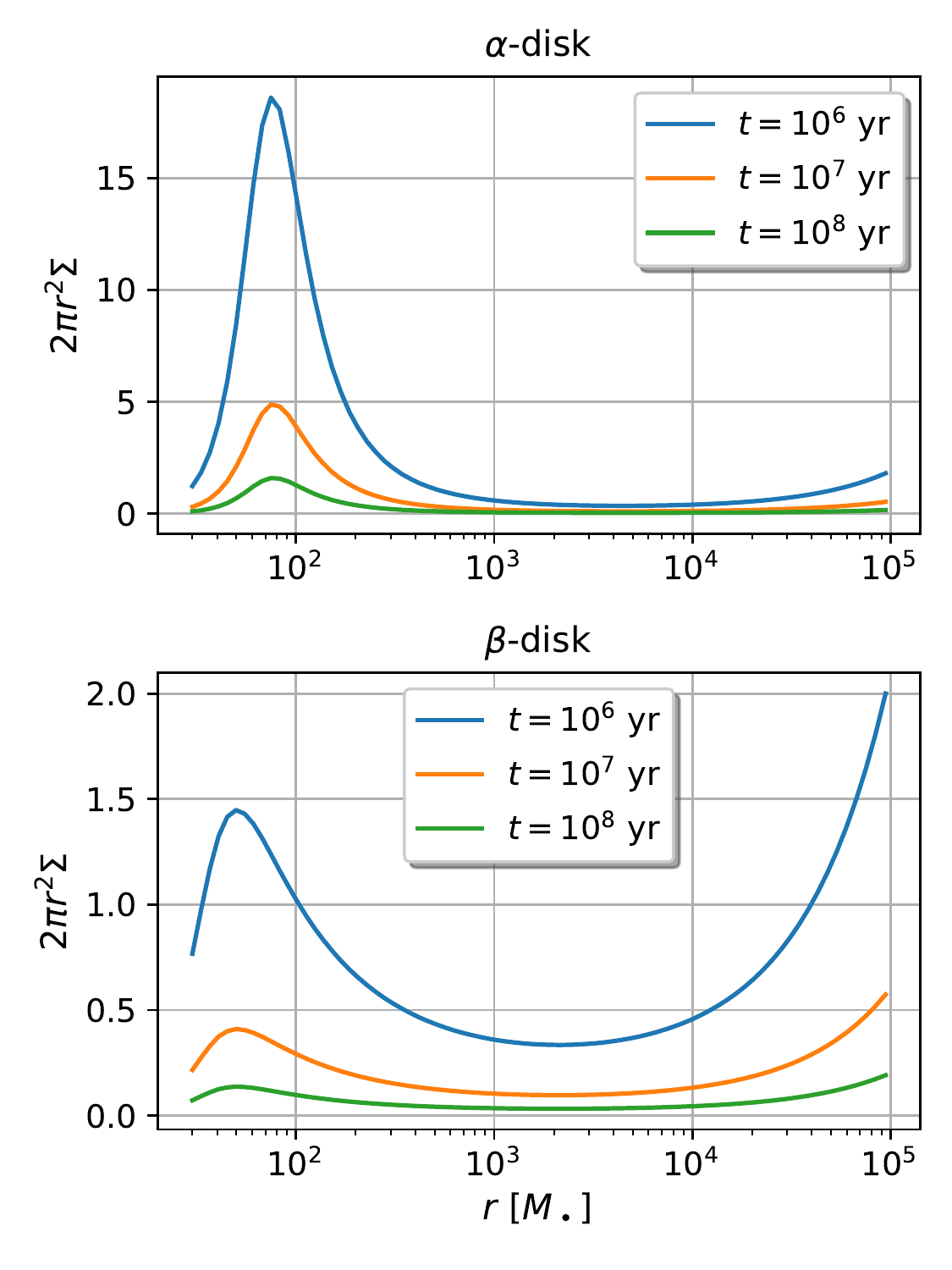}
\caption{\label{fig:surf_den} The surface number density $\Sigma(r)$ of disk-component sBHs at different times for the two different AGN disk models.}
\end{figure}

\subsection{Wet EMRI rates}
Considering that the accretion history of MBHs is likely episodic \cite{King2015,Schawinski2015} and the active phase of an MBH is in general much shorter that its quiet phase \cite{Soltan1982,Galametz2009,Macuga2019},
we simplify the duty cycle of an AGN as a long quiet phase of $T_0=5$ Gyr followed by
a short active phase of duration $T_{\rm disk}\in \{10^6, 10^7, 10^8\}$ yr. This simplied picture holds
if the cluster evolution driven by purely two-body scatterings is negligible during a quiet phase between two active episodes, so that adjacent episodes can be effectively glued together as we understand the hisotory of evolution.

For calculating the wet EMRI rate, we initialize the stellar cluster around a MBH and evolve the system for $T_0$ in the same way as in the previous section, then
turn on the accretion disk and initialize the cluster-component and disk-component distributions as
\be
\begin{aligned}
  f_i(t=0, E, \mu) &= 0.99 \int_0^1 f_i(t=T_0, E, R) \ dR\ , \\
  g_i(t=0, E) &= 0.01\int_0^1 f_i(t=T_0, E, R) \ dR \ ,
\end{aligned}
\ee
where the integrand $f_i(t=T_0, E, R)$ is the distribution function at the end of the quiet phase,
and the initial fraction $0.01$ is the typical disk aspect ratio \cite{Pan2021a},
the exact value of which does not matter
because the disk-component star densities are mostly determined by subsequent capture and migration processes.
With this initialization, we continue the evolution of $f_i(t,E,\mu)$ and $g_i(t,E)$ according to Eqs.(\ref{eq:FP_b},\ref{eq:FP_c}) for a duration $T_{\rm disk}$.

As a fiducial model of the wet channel, we again consider a fiducial model with a MBH and a stellar cluster same to in the previous section, and a fiducial $\alpha$-disk with lifetime $T_{\rm disk}=10^8$ yr (Fig.~\ref{fig:tmig}).
In the first 3 panels of Fig.~\ref{fig:disk}, we show the cluster-component distribution functions $f_i(E,\mu)$ at $t=10^7$ yr, where we see low-inclination ($\mu\rightarrow 1$)  orbiters have been captured into the disk, therefore the distribution function $f_i(E,\mu\rightarrow 1)$ is relatively lower. In the 4th panel, we show the $\mu$-integrated distribution functions $\bar f_i(E)=\int_{-1}^1 d\mu f_i(E,\mu)$
and the disk-component functions $\bar g_i(E)$ at $t=10^7$ yr. The disk component dominates over the cluster component in the vicinity of the MBH (large $E$), and the disk component peaks around $E\sim 10^5\sigma^2$ where the migration timescale peaks (Fig.~\ref{fig:tmig}). It is interesting to note that the disk-component distribution functions $g_i(E)$ are proportional to their
abundance $\delta_i$ with no dependence on different star masses $m_i$, simply because both the rate of stars captured onto the disk and the migration rate of stars along the disk are proportional to their masses $m_i$, and the local density
$g_i(E)$ is determined by the ratio of the two rates in the equilibrium state.

In Fig.~\ref{fig:Gamma_disk}, we show the wet EMRI rates for both sBHs and MGOs as functions of time.
Initally both of them are low because it takes sometime for sBHs and MGOs that are captured by the disk to migrate to
the MBH; and the two rates peaks around $t=3\times10^5$ yr and $t=10^6$ yr, respectively, because the migration timescale is inversely proportional to the star mass $m_i$. After the peak time, the rate of stars captured by the disk is in equilibrium with the corresponding EMRI rate, and both of them steadily decrease with time $\propto t^{-0.5}$, which is the typical behavior of diffusion processes with absorbing boundary conditions.
We find the wet EMRI rate of sBHs in the equilibrium state is higher than the dry EMRI rate by $\mathcal O(10^2-10^3)$ for the fiducial model, which is consistent with the result of the slow disk capture scenario in our previous work (Fig.~10 of Ref.~\cite{Pan2021a}), though the detailed time dependence does not match exactly because we approximated the effect of
density waves as inclination-independent and neglected
the contribution from dynamical friction in the previous work,
while we have used more refined prescription of star-disk interaction [Eq.~(\ref{eq:wave_friction})]
and self-consistent calculation of the disk capture rate [Eq.~(\ref{eq:FP_c})] in this work.
In the loss-cone channel, the EMRI rate of MGOs is largely suppressed compared to sBHs
because MGOs benefit little from the mass segregation (see Figs.~\ref{fig:lc} and \ref{fig:Gamma_lc}),
while in the disk channel, the EMRI rate of MGOs is lower than that of sBHs by a factor of $\sim m_{\rm sbh}/m_{\rm mgo}$ in the equilibrium state (Fig.~\ref{fig:Gamma_disk}).
As a result, we find the wet EMRI rate of MGOs in the equilibrium state
is higher than the dry rate by $\mathcal O(10^3-10^4)$ for the fiducial model.

As shown in Fig.~\ref{fig:tmig}, the migration timescale of a sBH in an AGN disk peaks around $r\sim 10^2 M_\bullet$,
where the GW emission start to dominate over other processes. As a result, a number of sBHs aggregate around this radius due to a traffic jam (4th panel of Fig.~\ref{fig:disk}). For clarity, we show the surface number density of
disk-component sBHs $\Sigma$
in Fig.~\ref{fig:surf_den} at different times. Compared with the $\beta$-disk,
more sBHs aggregate around $r\sim 10^2 M_\bullet$ in the fiducial $\alpha$-disk because of slower migration speed.
A similar result (with slightly higher surface number densities) was obtained in our previous work \cite{Pan2021c}
though the disk capture rate was not calculated from first principles.

Similar to the fiducial model, we initialize a stellar cluster of a MBH with mass in the range of $(10^5-10^7)\ M_\odot$,
evolve the MBH+cluster system for $T_0=5$ Gyr, then turn on an accretion disk around the MBH
and initialize the cluster-component and disk-component distributions, and continue the evolution
for $T_{\rm disk}$. The average wet EMRI rates of sBHs and MGOs
\be
\bar\Gamma_{\rm wet}=\frac{1}{T_{\rm disk}}\int_{0}^{T_{\rm disk}} \Gamma_{\rm emri, disk} dt\ ,
\ee
are summarized in Table~\ref{table}. For long disk lifetime $T_{\rm disk}\gtrsim 10^7$ yr, the wet EMRI rates of sBHs and MGOs are higher for heavier MBHs because more sBHs and MGOs are available around heavier MBHs.
For short disk lifetime $T_{\rm disk}=10^6$ yr, the trend reverses for $M_\bullet\gtrsim 10^6 M_\odot$
because longer migration timescale in AGN disks ($\gtrsim T_{\rm disk}$) around heavier MBHs
hinders the sBHs and MGOs captured by the AGN disk from migrating to the vicinity of the MBH and becoming EMRIs.
In comparison with the dry channel, we find the wet EMRI rate of sBHs is usually higher by $\mathcal O(10^1-10^3)$
and the wet EMRI rate of MGOs is usually higher by $\mathcal O(10^3-10^4)$.

\begin{table*}
\caption{\label{table}Average EMRI rates (Gyr$^{-1}$) of sBHs and MGOs per AGN, assuming sBHs and MGOs are produced from SN explosion with equal numbers. }
\resizebox{1.7\columnwidth}{!}{
\addtolength{\tabcolsep}{5pt}
\begin{tabular}{c c  | c   c  c }
AGN disk  & $M_\bullet/M_\odot$  &
$(\bar\Gamma_{\rm wet}^{\rm mgo},\bar\Gamma_{\rm wet}^{\rm sbh})_{T_{\rm disk}=10^6 {\rm yr}}$ &
$(\bar\Gamma_{\rm wet}^{\rm mgo},\bar\Gamma_{\rm wet}^{\rm sbh})_{T_{\rm disk}=10^7 {\rm yr}}$ &
$(\bar\Gamma_{\rm wet}^{\rm mgo},\bar\Gamma_{\rm wet}^{\rm sbh})_{T_{\rm disk}=10^8 {\rm yr}}$ \\
  \hline
  $\alpha$-disk& $1\times10^7$ & $(0.1\times10^3, 4.1\times10^4)$& $(1.2\times10^4,4.9\times10^4)$ & $(6.7\times10^3,1.7\times10^4)$   \\
  & $4\times10^6$ & $(0.7\times10^3,9.0\times10^4)$& $(1.2\times10^4,3.7\times10^4)$ & $(4.7\times10^3,1.2\times10^4)$  \\
  & $1\times10^6$ & $(1.0\times10^4,4.4\times10^4)$& $(7.4\times10^3,1.3\times10^4)$ & $(2.5\times10^3,4.3\times10^3)$\\
  & $4\times10^5$ & $(1.1\times10^4,1.7\times10^4)$& $(4.3\times10^3,4.5\times10^3)$ & $(1.5\times10^3,1.4\times10^3)$ \\
  & $1\times10^5$ & $(2.7\times10^3,1.6\times10^3)$& $(1.1\times10^3,0.6\times10^3)$ & $(0.5\times10^3,0.3\times10^3)$  \\
  \hline
  $\beta$-disk& $1\times10^7$ & $(4.6\times10^3, 9.3\times10^4)$& $(1.3\times10^4,5.1\times10^4)$ & $(6.7\times10^3,1.7\times10^4)$   \\
  & $4\times10^6$ & $(7.3\times10^3,1.1\times10^5)$& $(1.3\times10^4,3.8\times10^4)$ & $(4.7\times10^3,1.2\times10^4)$  \\
  & $1\times10^6$ & $(1.5\times10^4,4.8\times10^4)$& $(7.6\times10^3,1.4\times10^4)$ & $(2.5\times10^3,4.3\times10^3)$\\
  & $4\times10^5$ & $(1.2\times10^4,1.8\times10^4)$& $(4.4\times10^3,4.9\times10^3)$ & $(1.5\times10^3,1.4\times10^3)$ \\
  & $1\times10^5$ & $(3.7\times10^3,3.2\times10^3)$& $(1.1\times10^3,0.8\times10^3)$ & $(0.5\times10^3,0.3\times10^3)$  \\
  \hline
\end{tabular}
}
\end{table*}

\begin{table*}
\caption{\label{table2} Forecasted Total and LISA detectable (with SNR$\geq 20$) EMRI rates of sBHs and MGOs
in the redshift range $0<z<4.5$ assuming sBHs and MGOs are produced from SN explosions with equal numbers. For the wet channel, we have assumed a conservative AGN fraction $f_{\rm AGN}=1\%$ throughout the universe.}
\resizebox{2.0\columnwidth}{!}{
\addtolength{\tabcolsep}{5pt}
\begin{tabular}{c c c  c  | c  c }
Wet EMRIs &$f_\bullet$  & AGN disk  & $T_{\rm disk}$ [yr] &  Total rates of (MGOs, sBHs) [yr$^{-1}$] &  LISA detectable rates of (MGOs, sBHs) [yr$^{-1}$] \\
\hline
& $f_{\bullet,-0.3}$ & $\alpha$-disk & $10^6$ & (1900, 6400) & (50, 480) \\
                &   &  & $10^7$ & (1400, 2500) & (24, 130) \\
                &   &  & $10^8$ & (540, 860) & (10, 54) \\ \cline{2-6}
                &   & $\beta$-disk & $10^6$ & (2700, 8200) & (65, 530) \\
                                   &  &  & $10^7$ & (1400, 2600) & (24, 150) \\
                                   &  &  & $10^8$ & (540, 860) & (10, 54) \\ \cline{2-6}
\hline
& $f_{\bullet,+0.3}$ & $\alpha$-disk & $10^6$ & (110, 1000) & (3, 34) \\
                &   &  & $10^7$ & (180, 470) & (1, 10) \\
                &   &  & $10^8$ & (71, 160) & ($<1$, 3) \\ \cline{2-6}
                &   & $\beta$-disk & $10^6$ & (200, 1300) & (5, 38) \\
                                    &  &  & $10^7$ & (190, 500) & (2, 11) \\
                                    &  &  & $10^8$ & (71, 160) & ($<1$, 3) \\ \hline
Dry EMRIs& $f_\bullet$  &  &   &  Total rates of (MGOs, sBHs) [yr$^{-1}$] &  LISA detectable rates of (MGOs, sBHs) [yr$^{-1}$] \\
\hline
& $f_{\bullet,-0.3}$  &  &  & (79, 1300)  & (1, 120)  \\
& $f_{\bullet,+0.3}$  &  &  & (3, 130)  & ($<1$, 10)
\end{tabular}
}
\end{table*}

\section{Detection Prospects}\label{sec:detect}
In addition to the generic EMRI rates per MBH/AGN obtained in the previous two sections,
a few extra pieces of information are needed for calculating the LISA detectable EMRI rate: the mass function of MBHs $dN_\bullet/dM_\bullet$ and the fraction of MBHs living a stellar cusp which is supposed to be destroyed during
a MBH merger following a previous galaxy merger.

Following Ref.~\cite{Babak2017}, we consider two MBH mass functions in the range of $(10^4, 10^7) M_\odot$,
\be\label{eq:fbullet}
\begin{aligned}
  f_{\bullet,-0.3}: \frac{dN_\bullet}{d \log M_\bullet}
  &= 0.01 \left( \frac{M_\bullet}{3\times 10^6 M_\odot}\right)^{-0.3}\ {\rm Mpc}^{-3}\ ,\\
    f_{\bullet,+0.3}: \frac{dN_\bullet}{d \log M_\bullet}
  &= 0.002 \left( \frac{M_\bullet}{3\times 10^6 M_\odot}\right)^{+0.3}\ {\rm Mpc}^{-3}\ ,
\end{aligned}
\ee
where $f_{\bullet,-0.3}$ is an approximation to the mass function
in the model assuming MBHs are seeded by Pop-III stars and grow via
accretion and mergers \cite{Barausse2012},
and $f_{\bullet,+0.3}$ is purely a phenomenological model \cite{Gair2010}.

In the frame of observers on the earth, the differential dry and wet EMRI rates are written as
\be\label{eq:R2}
\begin{aligned}
  \frac{d^2\mathcal R_{\rm dry}}{dM_\bullet dz}
  &=\frac{1}{1+z} \frac{dN_\bullet}{dM_\bullet}\frac{dV_{\rm c}(z)}{dz} C_{\rm cusp}(M_\bullet,z)
  \bar\Gamma_{\rm dry}(M_\bullet; N_{p})\ , \\
  \frac{d^2\mathcal R_{\rm wet}}{dM_\bullet dz}
  &= \frac{f_{\rm AGN}}{1+z}\frac{dN_\bullet}{dM_\bullet}\frac{dV_{\rm c}(z)}{dz}C_{\rm cusp}(M_\bullet,z)
  \bar\Gamma_{\rm wet}(M_\bullet;\mathbb{M})\ ,
\end{aligned}
\ee
where $z$ is cosmological redshift,
$V_{\rm c}(z)$ is the comoving volume of the universe up to redshift $z$,
$C_{\rm cusp}(M_\bullet,z)$ is the fraction of MBHs embedded in a stellar cusp,
where we use the same $C_{\rm cusp}(M_\bullet,z)$ function as in \cite{Babak2017}
for models with MBH mass function $f_{\bullet, -0.3}$ and take $C_{\rm cusp}(M_\bullet,z)=1$
for models with MBH mass function $f_{\bullet, +0.3}$.
$\bar\Gamma_{\rm wet}(M_\bullet;\mathbb{M})$ is the average wet EMRI rate (Table~\ref{table})
and we conservatively take the AGN fraction as $f_{\rm AGN}=1\%$.

In consistent with Ref.~\cite{Babak2017}, we parametrize the average dry EMRI rate of sBHs as
\be\label{eq:dry_eff}
{\bar\Gamma_{\rm dry}^{\rm sbh}}(M_\bullet; N_{\rm p}) = C_{\rm dep}(M_\bullet; N_{p})
C_{\rm grow}(M_\bullet; N_{p})\Gamma_{\rm lc}^{\rm sbh}(M_\bullet)\ ,
\ee
where we take the number of prompt infalls per EMRI as $N_{\rm p}=10$ (Fig.~\ref{fig:Gamma_lc}),
the generic rate
\be\label{eq:dry}
\Gamma_{\rm lc}^{\rm sbh}(M_\bullet) = \Gamma_0 \left( \frac{M_\bullet}{10^6 M_\odot}\right)^{-0.19}\ ,
\ee
with $\Gamma_0 \in (30, 300)\ {\rm Gyr}^{-1}$ \cite{Babak2017,Pan2021b},
and two correction factors are correction from possible depletion of sBHs available $C_{\rm dep}(M_\bullet; N_{p})$ and correction capping the MBH mass growth via accreting sBHs (from both prompt infalls and EMRIs)
$C_{\rm grow}(M_\bullet; N_{p})$, respectively.
Though there is uncertainty of a factor of $\mathcal O(10)$ in the generic rate $\Gamma_{\rm lc}^{\rm sbh}$,
we will see that the uncertainty does not propagate to the average rate $\bar\Gamma_{\rm dry}^{\rm sbh}$ with the two corrections. The depletion correction is formulated as
\be
C_{\rm dep} = {\rm min.}\left\{\frac{T_{\rm dep}}{T_{\rm rlx}}, 1 \right\}\ ,
\ee
where $T_{\rm dep}$ is the depletion timescale of sBHs residing in the MBH influence sphere ($r_c=2M_\bullet/\sigma^2$)
\be
T_{\rm dep} = \frac{\Sigma m_{\rm sbh}}{(1+N_{\rm p})\Gamma_{\rm lc}^{\rm sbh} m_{\rm sbh}}\ ,
\ee
assuming the total mass of sBHs in the influence sphere is $\Sigma m_{\rm sbh} = 0.06 M_\bullet$\ ,
and the relaxation timescale at $r=r_c$ is
\be
T_{\rm rlx} = \left(\frac{\sigma}{20\ {\rm km/s}}\right) \left(\frac{r_c}{1\ {\rm pc}}\right)^2\ {\rm Gyr}\ .
\ee
The MBH growth correction comes from requiring the MBH mass grows no more than $1/e$ via accreting sBHs,
\be
C_{\rm grow} = {\rm min.}\left\{e^{-1}\frac{M_\bullet}{\Delta M_{\bullet}}, 1\right\}\ ,
\ee
with
\be
\Delta M_\bullet = m_{\rm sbh}(1+N_{\rm p}) C_{\rm dep}(M_\bullet, N_{\rm p})  \Gamma_{\rm lc}^{\rm sbh}(M_\bullet, N_{\rm p}) T_{\rm emri}(M_\bullet)\ ,
\ee
and
\be
T_{\rm emri} = \int dt \frac{dt}{dz} C_{\rm cusp}(M_\bullet, z)
\ee
is the total duration when a MBH lives in a stellar cusp.

With these two corrections, we find the average EMRI rate of sBHs is well fitted by
\be
\begin{aligned}
  &{\bar\Gamma_{\rm dry}^{\rm sbh}}(M_\bullet; N_{\rm p}=10) \\
  &= {\rm min.}\left\{ 26\left( \frac{M_\bullet}{10^5 M_\odot}\right), 30\left( \frac{M_\bullet}{10^6 M_\odot}\right)^{-0.19}\right\} \ {\rm Gyr}^{-1}\ .
\end{aligned}
\ee
with little dependence on the generic rate $\Gamma_0$ as long as it is higher than $30 \ {\rm Gyr}^{-1}$,
because the average rate is in fact determined by the number of sBHs available around MBHs
and the MBH growth limit via accreting sBHs.
For the average EMRI rate of MGOs, we simply take it as
$\bar\Gamma_{\rm dry}^{\rm mgo}(M_\bullet) \approx 1\ {\rm Gyr}^{-1}$
(Fig.~\ref{fig:Gamma_lc_all}).

With all the elements for calculating the differential EMRI rates ready [Eq.~(\ref{eq:R2})], we
calcuate the total EMRI rates of MGOs and sBHs from the two channels and the LISA detectable EMRI rates.
We first sample the EMRI sources according to the differential rates [Eq.~(\ref{eq:R2})], then
compute the EMRI waveform using the Augment Analytic Kludge \cite{Barack2004,Chua2015,Chua2017}
and the expected signal-to-ratio (SNR) by the LISA detector
(see all the source sampling and SNR computation details in the previous work \cite{Pan2021b}).
The forecast results are listed in Table~\ref{table2}. For the well-motivated MBH mass function
$f_{\bullet,-0.3}$, we expect LISA to detect  $\sim 1$ mass-gap EMRIs,
$\mathcal O(10^2)$ sBH EMRIs from the dry channel,
$\mathcal O(10-10^2)\times (f_{\rm AGN}/1\%)$ mass-gap EMRIs,
and $\mathcal O(10^2-10^3)\times (f_{\rm AGN}/1\%)$ sBH EMRIs from the wet channel
per year. For the less optimistic MBH mass function $f_{\bullet,+0.3}$, the expected
detection numbers are overall lower by a factor of $\mathcal O(10)$.

\section{Summary and Discussion}\label{sec:summary}
\subsection{Summary}
In the dry EMRI formation channel, the formation rate of mass-gap EMRIs is strongly
suppressed compared with EMRIs of sBHs, because sBHs are heavier and accumulate closer to
MBH due to the mass segregation effect and therefore easier to form EMRIs (Fig.~\ref{fig:Gamma_lc}).
In the wet channel, the EMRI formation turns out to be much more efficient than in the dry channel
because the capture of compact objects onto the accretion disk
and subsequent inward migration along the disk are highly efficient in transporting
compact objects  \cite{Pan2021a,Pan2021b}.
Both the capture rate onto to disk and the migration speed along the disk are linearly proportional to the
compact object mass, so that the formation rate of wet mass-gap EMRIs in the equilibrium state
is suppressed by a factor of $\sim m_{\rm mgo}/m_{\rm sbh}$ assuming their abundances are equal ($\delta_{\rm sbh}=\delta_{\rm mgo}$).
As a result, we find the wet EMRI rate of sBHs per AGN is higher
than the dry rate per MBH by $\mathcal O(10^1-10^3)$,
and the wet EMRI rate of MGOs per AGN is higher than the dry rate per MBH by $\mathcal O(10^3-10^4)$.
Accounting for the AGN fraction $f_{\rm AGN}=\mathcal O(10^{-2}-10^{-1})$, the wet channel turns out to
the dominant channel of mass-gap EMRI formation.
As for the LISA detection prospects,
we expect LISA to detect no more than $\sim 1$ dry mass-gap EMRIs,
and $\mathcal O(10-10^2)\times (f_{\rm AGN}/1\%)$ wet mass-gap EMRIs  per year
for the physically motivated MBH mass function $f_{\bullet,-0.3}$.
For the less optimistic MBH mass function $f_{\bullet,+0.3}$,
the expected detection numbers are lower by $\mathcal O(10)$  (Table~\ref{table2}).

\subsection{Discussion}
As shown above, the expected number of EMRI detections (denoted as $D_{\rm sbh}$ and $D_{\rm mgo}$)
are sensitive to the unknown MBH mass function, while the ratio $D_{\rm sbh}/D_{\rm mgo}$ is not, which
can be used as a more robust probe to the MGO abundance and origin.

If LISA detects $D_{\rm mgo}$ dry mass-gap EMRIs and $D_{\rm sbh}$ dry EMRIs of sBHs per year, we can infer the
relative abundance of MGOs and sBHs within nuclear stellar clusters as
\be\label{eq:D_ratio}
\frac{\delta_{\rm mgo}}{\delta_{\rm sbh}} \approx \frac{D_{\rm mgo}/D_{\rm sbh}}{R^{\rm mgo}_{\rm sbh}}\ ,
\ee
where $R^{\rm mgo}_{\rm sbh}\approx 1/120$ is the ratio of expected detection numbers of two different EMRIs
assuming MGOs and sBHs are of the same abundance (Table~\ref{table2}).
In a similar way making use of Eq.~(\ref{eq:D_ratio}), one can again infer the
relative abundance of MGOs $\delta_{\rm mgo}/\delta_{\rm sbh}$ from detections of wet EMRIs,
where $R^{\rm mgo}_{\rm sbh} =  (1/10-1/5)$ (Table~\ref{table2}) varies little over all different model parameters for the parameter space we considered. The inferred relative abundance $\delta_{\rm mgo}/\delta_{\rm sbh}$ can be used to constrain the SN explosion mechanisms, where the delayed SN explosion mechanism predicts $\delta_{\rm mgo}/\delta_{\rm sbh}\rightarrow 1$ while the rapid explosion mechanism predicts $\delta_{\rm mgo}/\delta_{\rm sbh}\rightarrow 0$ \cite{Fryer2012}.

An excess of mass-gap EMRI detection by LISA is a signature of MGOs of exotic origins (e.g., PBHs \cite{Carr2019,Clesse2020,Jedamzik2021,Guo2019,Barsanti2021}). If these MGOs are of primordial origin, their abundance around MBHs may be further used to constrain the mass fraction of
PBHs in dark matter (DM) $f_{\rm mgo}:=\Omega_{\rm mgo}/\Omega_{\rm DM}$. This constraint sensitively depends on the DM distribution around MBHs, which is poorly understood theoretically. We consider two extremal cases: (1)
DM around MBHs traces baryons with the DM to baryon ratio $\Omega_{\rm DM}/\Omega_{\rm B}$;
(2) the DM density around MBHs follows the NFW profile \cite{NFW1996}.

In case (1), the abundance of MGOs relative to stars is
\be
  \frac{\delta_{\rm mgo}}{\delta_{\rm star}}
  = \frac{f_{\rm mgo}\Omega_{\rm DM}}{(1-f_{\rm gas})\Omega_{\rm B}}\frac{m_{\rm star}}{m_{\rm mgo}}\ ,\\
\ee
where $f_{\rm gas}$ is the mass fraction of gas in baryons
and $f_{\rm mgo}$ is formulated as
\be\label{eq:abundance}
\begin{aligned}
  f_{\rm mgo}
  &= (1-f_{\rm gas})\frac{\delta_{\rm sbh}}{\delta_{\rm star}} \frac{D_{\rm mgo}/D_{\rm sbh}}{R^{\rm mgo}_{\rm sbh}}
  \frac{\Omega_{\rm B}}{\Omega_{\rm DM}}\frac{m_{\rm mgo}}{m_{\rm star}} \\
  &= 6\times10^{-4}(1-f_{\rm gas})
  \frac{\delta_{\rm sbh}/\delta_{\rm star}}{10^{-3}} \frac{D_{\rm mgo}/D_{\rm sbh}}{R^{\rm mgo}_{\rm sbh}}
  \frac{\Omega_{\rm B}/\Omega_{\rm DM}}{0.2}\frac{m_{\rm mgo}/m_{\rm star}}{3}
\end{aligned}
\ee
where we have used Eq.(\ref{eq:D_ratio}).

In case (2), the DM abundance around MBHs is usually much lower, with the total DM mass with the MBH influence radius
$M_{\rm DM}(<r_c)\approx 0.3\% M_\bullet$ (see Appendix~\ref{app_c} for details). For comparison, the total star mass
is $\Sigma m_{\rm star}(<r_c)\approx 2 M_\bullet$ and the total mass of astrophysical MGOs is
$\Sigma m_{\rm mgo}(r<r_c)\approx 0.6\% M_\bullet \times (\delta_{\rm mgo}/10^{-3})$, i.e.,
the DM abundance around MBHs is comparable with that of astrophysical MGOs.
In this case, it is unlikely to observe  excess of mass-gap EMRIs. Therefore,
an excess of mass-gap EMRI detection by LISA would disfavor the NFW distribution of MGOs as DM.

Many other proposals to explain GW190814-like events involve hierachical mergers, e.g., in young stellar clusters \cite{Rastello2020}, triple systems \cite{Lu2021}, AGN disks \cite{Tagawa2021b}, etc. If the abundance of MGO production can be realibly estimated in these scenarios, the rate of mass gap EMRIs may also be used to test these models. For example, as young stellar clusters are not expected to host massive black holes, if they are the only places that MGOs are produced, we should expect the mass-gap EMRI rate to be minimized.

One working assumption we used is no mass change of MGOs in AGN disks.
As shown in Fig.~\ref{fig:tmig}, the typical migration timescale of MGOs is $\sim 10^6$ yr,
which is much shorter than the Salpeter timescale $5\times10^7$ yr (mass doubling timescale with the Eddington accretion rate). But the accretion rate of MGOs/sBHs in AGN disks is uncertain \cite{Pan2021c,Tagawa2022}.
If gas accretion onto MGOs was highly super-Eddington and largely increased their masses,
identifying the distorted mass gap is less straightforward, and the commonly used
techniques of searching for mass-gap features
in the mass spectrum of LIGO/Virgo events \cite{LVC2021} would also be valuable for our purpose.

\acknowledgements
Z. P. and H. Y. are supported by the Natural Sciences and
Engineering Research Council of Canada and in part by
Perimeter Institute for Theoretical Physics. Research at
Perimeter Institute is supported in part by the Government of
Canada through the Department of Innovation, Science and
Economic Development Canada and by the Province of Ontario through the Ministry of Colleges and Universities.

\appendix
\section{Diffusion and advection coefficients in the Fokker-Planck equation (\ref{eq:FP})}\label{apa}
In Refs.~\cite{Shapiro1978,Cohn1978,Cohn1979}, the diffusion and the advection coefficients of a single-species cluster
have been derived in detail. Following Refs.~\cite{Binney1987, Bar-Or2016},  we extend the them to multi-species cases. We first define auxiliary functions:
\be
\begin{aligned}
  F_0^{(j)}(E,r) &= (4\pi m_j)^2 \ln\Lambda \int_{-\infty}^E dE' \bar f_j(E') \ ,\\
  F_n^{(j)}(E,r) &= (4\pi m_j)^2 \ln\Lambda \int_E^{\phi(r)} dE' \left(\frac{\phi-E'}{\phi-E}\right)^{n/2} \bar f_j(E') \ ,
\end{aligned}
\ee
where $n\geq 1$, $j$ is the index labelling different star species, $\ln\Lambda$ the Coulomb's logarithm which we take as $\ln\Lambda=10$, and
\be
\bar f_j(E) := \int_0^1 f_j(E,R) dR\ .
\ee
With these auxiliary functions, the coefficients are written as
\be
\begin{aligned}
  D_{EE}^{(i)} &= \sum_j\frac{2}{3P}\int_{r_-}^{r_+} \frac{dr}{v_r} v^2(F_0^{(j)}+F_3^{(j)})
  \ ,\\
  D_{E}^{(i)} &= \sum_j-\frac{2}{P}\int_{r_-}^{r_+} \frac{dr}{v_r} F_1^{(j)}\times \frac{m_i}{m_j}\ ,\\
  D_{ER}^{(i)} &= \sum_j\frac{4}{3P}R\int_{r_-}^{r_+} \frac{dr}{v_r} \left(\frac{v^2}{v_c^2}-1\right)(F_0^{(j)}+F_3^{(j)})\ ,\\
  D_{RR}^{(i)} &= \sum_j\frac{4}{3 P}\frac{R}{J_c^2}\int_{r_-}^{r_+} \frac{dr}{v_r}
  \Bigg\{2\frac{r^2}{v^2} \left[v_t^2\left(\frac{v^2}{v_c^2}-1\right)^2 +v_r^2\right]F_0^{(j)} \\
  &+ 3\frac{r^2}{v^2}v_r^2F_1^{(j)}
  +\frac{r^2}{v^2}\left[2v_t^2\left(\frac{v^2}{v_c^2}-1\right)^2 -v_r^2 \right]F_3^{(j)}\Bigg\}\ ,\\
  D_R^{(i)}&= \sum_j-\frac{4}{P} \frac{R}{v_c^2} \int_{r_-}^{r_+}  \frac{dr}{v_r} \left(1-\frac{v_c^2}{v^2}\right) F_1^{(j)}
  \times\frac{m_i}{m_j}\ ,
\end{aligned}
\ee
where $i$ (similar to $j$) is also the star species index,  $v_t = J/r$ is the tangential velocity, and $v_c(E)$ is the velocity of a circular orbit with energy $E$.

\section{Diffusion and advection coefficients in the Fokker-Planck equation (\ref{eq:FP_b})}\label{apb}
Following Refs.~\cite{Cohn1978,Cohn1979,Binney1987,Bar-Or2016}, we derive the diffusion and advection coefficients in the Fokker-Planck equation (\ref{eq:FP_b}) in this section.
Consider a star $m$ with orbital energy $E$ and velocity $\bm{v}$ changes its energy by $\Delta E$ and velocity by $\Delta \bm v$ due to a scattering with a field star $m_a$. In the following orthonormal basis,
\be
\begin{aligned}
  \hbm{v} &= \bm v/v\ , \\
  \hbm{J} &= \bm J/J = \bm r\times \bm v/J\ ,\\
  \hbm{w} &= \bm v\times \bm J/|\bm v\times \bm J|\ ,
\end{aligned}
\ee
the velocity change is written as
\be
\Delta \bm v = \Delta v_\parallel \hbm{v} + \Delta \bm v_\perp
 =\Delta v_\parallel \hbm{v} + \Delta v_J \hbm{J} + \Delta v_w \hbm{w}\ .
\ee
As a result, the changes in energy and angular momentum are
\be
\Delta E =  -\frac{1}{2}(\Delta v_\parallel)^2
-\frac{1}{2} (\Delta v_\perp)^2 -v \Delta v_\parallel\ ,
\ee
and
\be
\begin{aligned}
  \Delta\bm J
  &= \bm{r}\times \Delta\bm{v} \\
  &=J\pr{\frac{\Delta v_\parallel}{v}-\frac{v_r}{v_t}\frac{\Delta v_w}{v}}\hbm{J}
  +J\frac{\Delta v_J}{v}\pr{\frac{v_r}{v_t}\hbm w-\hbm v} \\
  &:= \Delta \bm J_\parallel + \Delta\bm{J}_\perp\ ,
\end{aligned}
\ee
with $\Delta J_\perp = J(\Delta v_J/v)\sqrt{1+(v_r/v_t)^2}= J(\Delta v_J/v_t) =r \Delta v_J$ and
\be
\begin{aligned}
  \Delta J := |\bm J + \Delta\bm J| - J = J\frac{\Delta v_\parallel}{v}-rv_r\frac{\Delta v_w}{v} + \frac{1}{2}\frac{r^2(\Delta v_J)^2}{J}\ .
\end{aligned}
\ee
For $\mu = \cos\iota = J_z/J$ with $J_z$ the $z$-component angular momentum, its change is
\be
\Delta \mu = \frac{\Delta J_z}{J}-\mu\frac{\Delta J}{J}
= \frac{\Delta v_J}{v}\pr{\frac{v_r}{v_t}\hbm{w}-\hbm{v}}\cdot\hbm{z}
-\frac{\mu}{2}\frac{r^2(\Delta v_J)^2}{J^2}\ .
\ee

Defining  $\braket{\Delta X}_t:=\frac{\braket{\Delta X}}{\Delta t}|_{\Delta t\rightarrow 0}$
and  $\braket{\Delta X\Delta Y}_t:=\frac{\braket{\Delta X\Delta Y}}{\Delta t}|_{\Delta t\rightarrow 0}$
(where $\braket{}$ is ensemble average over scatterings a star has experienced),
it is straightforward to see
\be
\begin{aligned}
  \braket{\Delta E}_t
  &= -\frac{1}{2}\braket{(\Delta v_\parallel)^2}_t
  -\frac{1}{2}\braket{(\Delta v_\perp)^2}_t -v \braket{\Delta v_\parallel}_t\ ,\\
  \braket{(\Delta E)^2}_t
  &=  v^2 \braket{(\Delta v_\parallel)^2}_t\ , \\
  \braket{\Delta\mu}_t
  &= \braket{\frac{\Delta v_J}{v}}_t\pr{\frac{v_r}{v_t}\hbm{w}-\hbm{v}}\cdot\hbm{z}-\frac{\mu}{2}\frac{r^2}{J^2}\braket{(\Delta v_J)^2}_t\ ,\\
  \braket{(\Delta\mu)^2}_t
  &= \braket{   \pr{\frac{\Delta v_J}{v}\pr{\frac{v_r}{v_t}\hbm{w}-\hbm{v}}\cdot\hbm{z}}^2 }_t\\
  &=\braket{ \frac{(\Delta v_J)^2}{v^2} \pr{\frac{v_r^2}{v_t^2}+1}\frac{\sin^2\iota}{2} }_t \\
  &=\frac{1-\mu^2}{2}\frac{r^2}{J^2}\braket{(\Delta v_J)^2}_t\ ,\\
  \braket{\Delta E\Delta\mu}_t
  &= -\pr{\frac{v_r}{v_t}\hbm{w}-\hbm{v}}\cdot\hbm{z}\braket{\Delta v_\parallel\Delta v_J}_t\ ,
\end{aligned}
\ee
accurate to quadractic order in $\Delta v$.

In the case of spherical symmetry, coefficients $\braket{\Delta \bm v}_t$ and $\braket{\Delta \bm v\Delta \bm v}_t$
has been derived by \citet{Binney1987} (assuming the field stars $m_a$ are symmetrically distributed in the azimuthal direction in the rest frame of particle $m$) as
\be
\begin{aligned}
  \braket{\Delta v_\parallel}_t
  &= -\kappa \frac{m+m_a}{m_a} \int_0^v  dv_a \frac{v_a^2}{v^2}  f_a(v_a)\ ,\\
  \braket{(\Delta v_\parallel)^2}_t
   &= \frac{2}{3}\kappa\pr{\int_0^v dv_a\frac{v_a^4}{v^3} f_a(v_a)+\int_v^\infty dv_a v_a f_a(v_a)} \\
   \braket{(\Delta v_\perp)^2}_t
   &= \frac{2}{3}\kappa\pr{\int_0^v dv_a\pr{\frac{3v_a^2}{v}-\frac{v_a^4}{v^3}} f_a(v_a)+2\int_v^\infty dv_a v_af_a(v_a)}
\end{aligned}
\ee
$\braket{\Delta \bm v_\perp}_t = 0$ and $\braket{\Delta v_\parallel\Delta v_J}_t=0$, with $\kappa = (4\pi m_a)^2 \ln\Lambda$.
It is straightforward to extend the above results to the non-spherical symmetry case,
with the replacement $f_a(v_a)\rightarrow \bar f_a(v_a,\theta)$, where
\be
\bar f_a(v_a,\theta):=\frac{1}{2\pi}\int_0^{2\pi} d\eta f_a(v_a, \mu=\sin\theta\cos\eta) d\eta\ ,
\ee
and $\theta$ is the polar angle w.r.t. the $z$-axis.

With $\braket{\Delta \bm v}_t$ and $\braket{\Delta \bm v\Delta \bm v}_t$ ready, the derivation of coefficients in
the Fokker-Planck equation (\ref{eq:FP_b}) is parallel to the previous section and we outline it as follows.
Define functions
\be
\begin{aligned}
  F_0^{(j)}(E,  \theta) &= (4\pi m_j)^2 \ln\Lambda \int_{-\infty}^E dE' \bar f_j(E', \theta), \\
  F_n^{(j)}(E, r, \theta) &= (4\pi m_j)^2 \ln\Lambda \int_E^{\phi(r)} dE' \bar f_j(E', \theta)\pr{\frac{\phi-E'}{\phi-E}}^{n/2},
\end{aligned}
\ee
with $n\geq 1$ and
\be
\bar f_j(E,\theta):=\frac{1}{2\pi}\int_0^{2\pi} d\eta f_j(E, \mu=\sin\theta\cos\eta) d\eta\ .
\ee
With these auxiliary functions, the local diffusion/advection coefficients
$\hat D(E,R,r,\theta)$ (which depend on energy $E$, normalized angular momentum $R$,
and spatial coordinates $r$ and $\theta$) are
\be
\begin{aligned}
  \hat D_{EE}^{(i)} &=\sum_j \frac{1}{3}v^2(F_0^{(j)}+F_3^{(j)})\ , \\
  \hat D_{E}^{(i)}  &=\sum_j -F_1^{(j)}\times\frac{m_i}{m_j}\ , \\
  D_{\mu\mu}^{(i)} &= \sum_j\frac{1-\mu^2}{4}\frac{1}{3}  v_t^{-2}(2F_0^{(j)}+3F_2^{(j)}-F_4^{(j)})\ ,\\
  D_{\mu}^{(i)} &= \sum_j\frac{1-\mu^2}{4}\frac{1}{3}  \pder{\mu} v_t^{-2}(2F_0^{(j)}+3F_2^{(j)}-F_4^{(j)})\ ,\\
  D_{\mu E}^{(i)} &= 0\ .
\end{aligned}
\ee

In general, the orbital coordinates $(r, \theta)$  of a star orbit with semi-major axis $a(E)$ and eccentricity $e$
are specified by
\be
r = \frac{a(1-e^2)}{1+e\cos\psi'}\ ,\quad \cos\theta = \cos(\psi'+\phi)\sin\iota\ ,
\ee
with $\phi\in[0,2\pi]$ specifying the pericenter location in the azimuthal direction, and $\psi'\in[0,2\pi]$ is the orbital phase w.r.t the pericenter.
With orbit average, we obtain
\be
\bar D(E,R,\mu, \phi) = \frac{2}{P}\int_{r_-}^{r_+} \frac{dr}{v_r} \hat D(E,R,r,\theta)\ .
\ee
After ensemble average over $R$ and $\phi$, we arrive at the final form $D(E,\mu)=\braket{\bar D(E,R,\mu, \phi)}_{R,\phi}$.

In fact, we find circular orbits is a good approximation in calculating the coefficients.
For circular orbits, the orbital equation is simplified as
\be
r = r_c(E),\quad \cos\theta = \cos\psi'\sin\iota
\ee
and the diffusion/advection coefficients arising from scatterings with the cluster-component stars
\be\label{eq:D_cls}
\begin{aligned}
   D_{EE}^{(i)} &= \sum_j\frac{1}{3\pi}  \int_0^\pi d\psi' v^2(F_0^{(j)}+F_3^{(j)})\ , \\
   D_{E}^{(i)} &= \sum_j-\frac{1}{\pi} \int_0^\pi d\psi'  F_1^{(j)} \times \frac{m_i}{m_j} \ ,\\
  D_{\mu\mu}^{(i)} &= \sum_j\frac{1-\mu^2}{4}\frac{1}{3\pi} \int_0^\pi d\psi'  v_t^{-2}(2F_0^{(j)}+3F_2^{(j)}-F_4^{(j)}) \ ,\\
  D_{\mu}^{(i)} &= \sum_j\frac{1-\mu^2}{4}\frac{1}{3\pi} \pder{\mu}\int_0^\pi d\psi'  v_t^{-2}(2F_0^{(j)}+3F_2^{(j)}-F_4^{(j)}) \ ,\\
  D_{\mu E}^{(i)} &= 0\ .
\end{aligned}
\ee

With the same approximation, the diffusion/advection coefficients arising from scatterings
with the disk-component stars are
\be\label{eq:D_dsk}
\begin{aligned}
   D_{EE}^{(i)} &= \sum_j\frac{1}{3}\epsilon  v^2(G_0^{(j)}+G_3^{(j)})\ ,  \\
   D_{E}^{(i)} &=  \sum_j-\epsilon G_1^{(j)} \times \frac{m_i}{m_j}\ ,\\
  D_{\mu\mu}^{(i)} &
  = \sum_j\frac{1-\mu^2}{4}\frac{1}{3} \epsilon v_t^{-2}(2G_0^{(j)}+3G_2^{(j)}-G_4^{(j)})\ ,\\
  D_{\mu}^{(i)} &
  =\sum_j\frac{1-\mu^2}{4}\frac{1}{3} \pder{\mu} \epsilon v_t^{-2}(2G_0^{(j)}+3G_2^{(j)}-G_4^{(j)})\ ,\\
  D_{\mu E}^{(i)} &= 0\ .
\end{aligned}
\ee
where $\epsilon(h,\iota) = {\rm min}\{1, \frac{2}{\pi}\arcsin(\frac{h}{\iota}) \}\approx  \frac{2}{\pi}\arcsin(\frac{h}{h+\iota})$ is fraction of the orbit lying inside the disk component,
and
\be
\begin{aligned}
  G_0^{(j)}(E) &=  (4\pi m_j)^2\ln\Lambda \int_{-\infty}^E dE' g_j(E'), \\
  G_n^{(j)}(E, r) &=  (4\pi m_j)^2\ln\Lambda \int_E^{\phi(r)} dE' g_j(E')\pr{\frac{\phi-E'}{\phi-E}}^{n/2}\ ,
\end{aligned}
\ee
with $n\geq 1$.

\section{DM NFW profile}\label{app_c}
If DM density around a MBH follows the NFW profile \cite{NFW1996}
\be
\rho_{\rm DM}(r) = \frac{\rho_s}{\frac{r}{R_s} \left(1+ \frac{r}{R_s}\right)^2},
\ee
the total DM mass within radius $r$ is written as
$M_{\rm DM}(<r) = 4\pi\rho_s R_s^3 \mathcal G(c)$, with concentration $c:=r/R_s$ and $\mathcal G(c) = \ln(1+c)-c/(1+c)$,
where $\rho_s$ and $R_s$ are the characteristic density and radius, respectively.
For relating $\rho_s$ and $R_s$ to the MBH mass $M_\bullet$, we need the aid of a commonly used cutoff radius
within which the average DM density is 200 times the critical density of the universe $\rho_{\rm crit}$, i.e.,
\be
M_{\rm 200} = 4\pi\rho_s R_s^3 \mathcal G(c_{200}), \quad \frac{\rho_s}{\rho_{\rm crit}} = \frac{200}{3}\frac{c_{200}^3}{\mathcal G(c_{200}) } \ ,
\ee
As found in Ref.~\cite{Ferrarese2002}, $M_{200}$ and $M_\bullet$ are correlated with
\be
\frac{M_\bullet}{10^7 M_\odot} \approx \left(\frac{M_{200}}{10^{12} M_\odot} \right)^{1.65}\ .
\ee
For low redshift, the concentration $c_{200}$ has a weak dependence on the mass $M_{200}$ \cite{Dutton2014}
\be
\log_{10} c_{200} = 0.905 - 0.101 \log_{10}(M_{200}/10^{12}h^{-1} M_\odot)\ .
\ee
Combining the above three equations, we find the total mass of DM within the influence radius $r_c$
is $M_{\rm DM}(<r_c)\approx 0.3\% M_\bullet$ for $M_\bullet\in (10^5, 10^7) M_\odot$.

\bibliography{ms}
\end{document}